\documentclass[10pt,journal,tnecolumn]{IEEEtran}
\IEEEoverridecommandlockouts
\usepackage{cite}
\usepackage[bookmarks,colorlinks]{hyperref} 
\hypersetup{colorlinks,citecolor= red,filecolor= blue,linkcolor= blue,urlcolor=blue}
\usepackage{amsmath,amssymb,amsfonts}
\usepackage{xfrac}
\usepackage{graphicx}
\usepackage{subfig}
\usepackage{float}
\usepackage{textcomp}
\usepackage{xcolor}
\usepackage{balance}
\usepackage{winsnotation}
\usepackage{algorithm}
\usepackage{algorithmicx,tikz}
\usepackage{algcompatible}
\usepackage{tikz}
\usepackage[utf8]{inputenc}
\usepackage[english]{babel}
\usepackage{amsthm}

\usepackage[font=small,labelfont=bf]{caption}
\def\BibTeX{{\rm B\kern-.05em{\sc i\kern-.025em b}\kern-.08em
    T\kern-.1667em\lower.7ex\hbox{E}\kern-.125emX}}

\newcommand{\bd}{\begin{description}}
\newcommand{\ed}{\end{description}}
\newcommand{\be}{\begin{enumerate}}
\newcommand{\ee}{\end{enumerate}}
\newcommand{\bi}{\begin{itemize}}
\newcommand{\ei}{\end{itemize}}
\newcommand{\bl}{\begin{list}}
\newcommand{\el}{\end{list}}
\newcommand{\bt}{\begin{tabbing}}
\newcommand{\et}{\end{tabbing}}

\definecolor{BLUE}{rgb}{0,0,1}

\usepackage{acronym}  
\acrodef{dl}[DL]{deep learning}
\acrodef{dnn}[DNN]{deep neural network}
\acrodef{sic}[SIC]{successive interference cancellation}
\acrodef{nn}[NN]{neural network}
\acrodef{pdf}[PDF]{probability density function}
\acrodef{pmf}[PMF]{probability math function}
\acrodef{sto}[STO]{symbol timing offset}
\acrodef{isi}[ISI]{intersymbol interference}
\acrodef{awgn}[AWGN]{additive white Gaussian noise}
\acrodef{iot}[IoT]{Internet of Things}
\acrodef{cdma}[CDMA]{code division multiple access}
\acrodef{bs}[BS]{base station}
\acrodef{bpsk}[BPSK]{binary phase-shift keying}
\acrodef{dsss}[DS-SS]{direct-sequence spread spectrum}
\acrodef{sssr}[SSSR]{simultaneous sparse signal reconstruction}
\acrodef{tdd}[TDD]{time-delay diversity}
\acrodef{tddssa}[TDD-SSA]{time-delay diversity sparse signal approximation}
\acrodef{kkt}[KKT]{Karush-Kuhn-Tucker}
\acrodef{map}[MAP]{maximum a posteriori probability}
\acrodef{somp}[S-OMP]{Simultaneous Orthogonal Matching Pursuits}
\acrodef{ls}[LS]{least square}
\acrodef{mud}[MUD]{multiuser detection}
\acrodef{mse}[MSE]{mean squared error}
\acrodef{ber}[BER]{bit error rate}
\acrodef{wcss}[WCSS]{within-cluster sum of squares}
\acrodef{sb}[SB]{Symbol-Based}
\acrodef{rd}[RD]{ridge regression}
\acrodef{ssr}[SSR]{sparse signal reconstruction}
\acrodef{iid}[i.i.d.]{independent and identically distributed}
\acrodef{ls}[LS]{least-squares}
\acrodef{mse}[MSE]{mean-squared error}
\acrodef{2mc}[2MC]{2-mean clustering}
\acrodef{sae}[SAe]{sparsity-aware}
\acrodef{wcss}[WCSS]{least within-cluster sum of squares}
\acrodef{pb}[PB]{Packet-Based}
\acrodef{pmf}[PMF]{probability mass function}
\acrodef{cv}[CV]{cross-validation}
\acrodef{mmse}[MMSE]{minimum mean squared error}
\acrodef{snr}[SNR]{signal-to-noise ratio}
\acrodef{cvpl}[CVPL]{cross-validated partial likelihood}
\acrodef{sdl}[SDL]{sparse dictionary learning}
\acrodef{mod}[MOD]{method of optimal directions}
\acrodef{stls}[S-TLS]{sparse-total least square}
\acrodef{cv}[CV]{cross validation}
\acrodef{mlr}[MLR]{maximum likelihood ratio}
\acrodef{gcv}[GCV]{generalized cross validation}
\acrodef{cdf}[CDF]{cumulative distribution function}
\acrodef{pls}[P-LS]{penalized-LS}
\acrodef{ldpc}[LDPC]{low-density parity-check}
\acrodef{dc}[DC]{Decision Combining}
\acrodef{ec}[EC]{Estimate Combining}
\acrodef{roc}[ROC]{receiver operating characteristic}
\acrodef{mtc}[MTC]{machine-type communications}
\acrodef{ma}[MA]{multiple access}
\acrodef{mac}[MAC]{media access control}
\acrodef{phy}[PHY]{physical}
\acrodef{ra}[RA]{random access}
\acrodef{fcc}[FCC]{fading channel coefficients}
\acrodef{cfo}[CFO]{carrier frequency offset}
\acrodef{ble}[BLE]{Bluetooth low-energy}
\acrodef{rfid}[RFID]{radio frequency identification}
\acrodef{csma}[CSMA]{carrier sensing multiple access}
\acrodef{ca}[CA]{collision avoidance}
\acrodef{lte}[LTE]{Long-Term Evolution}
\acrodef{rfid}[RFID]{radio frequency identification}
\acrodef{fsa}[FSA]{frame slotted ALOHA }
\acrodef{lpwa}[LPWA]{low-power wide area}
\acrodef{rftdm}[R-FTDM]{random frequency-time division multiplexing}
\acrodef{rpma}[RPMA]{random phase multiple access}
\acrodef{cdma}[CDMA]{code division multiple access}
\acrodef{lbt}[LBT]{listen-before-talk}
\acrodef{ss}[SS]{spread spectrum}
\acrodef{bch}[BCH]{Bose–Chaudhuri–Hocquenghem}
\acrodef{ap}[AP]{Access Point}
\acrodef{3gpp}[3GPP]{3rd Generation Partnership Project}
\acrodef{nb}[NB]{narrowband}
\acrodef{fdma}[FDMA]{frequency division multiple access}
\acrodef{rach}[RACH]{Random Access Channel}
\acrodef{fa}[FA]{fixed assignment}
\acrodef{tdma}[TDMA]{time-division multiple access}
\acrodef{ds}[DS]{direct sequence}
\acrodef{dsa}[DSA]{device sparsity-aware}
\acrodef{pdsa}[PDSA]{packet-device sparsity-aware}
\acrodef{pts}[PTS]{packet transmission state}
\acrodef{mld}[MLD]{maximum likelihood decoding}
\acrodef{cb}[CB]{contention-based}
\acrodef{css}[CSS]{chirp spread spectrum}
\acrodef{mf}[MF]{matched filter}
\acrodef{ltem}[LTE-M]{Long Term Evolution for Machines}
\acrodef{cp}[CP]{carrier phase}
\acrodef{per}[PER]{packet error rate}
\acrodef{dcd}[DCD]{differentially coherent decorrelation}
\acrodef{ca}[CA]{code-aided}
\acrodef{adam}[Adam]{adaptive moment estimation}
\acrodef{mimo}[MIMO]{multiple-input multiple-output}
\acrodef{nl}[NC]{nulling and cancelling}
\acrodef{spi}[SD-IRS]{sphere decoding with increasing radius search}

\makeatletter
\def\BState{\State\hskip-\ALG@thistlm}
\makeatother

\DeclareMathAlphabet{\pazocal}{OMS}{zplm}{m}{n}

\newtheorem{innercustomgeneric}{\customgenericname}
\providecommand{\customgenericname}{}
\newcommand{\newcustomtheorem}[2]{%
  \newenvironment{#1}[1]
  {%
   \renewcommand\customgenericname{#2}%
   \renewcommand\theinnercustomgeneric{##1}%
   \innercustomgeneric
  }
  {\endinnercustomgeneric}
}

\newtheorem{theorem}{Theorem}[section]

\newcustomtheorem{customthm}{Theorem}
\newcustomtheorem{customlemma}{Lemma}

\usepackage[yyyymmdd,hhmmss]{datetime} 
\newdateformat{monthyeardate}{\monthname[\THEMONTH] \THEDAY, \THEYEAR} 

\begin{document}

\renewcommand{\figurename}{Fig.}
{
	\twocolumn
	
	

	
\title{Deep Learning Based Sphere Decoding}


\author{
    Mostafa~Mohammadkarimi,~\IEEEmembership{Member,~IEEE},
     Mehrtash Mehrabi,~\IEEEmembership{Student~Member,~IEEE},\\
    Masoud Ardakani,~\IEEEmembership{Senior~Member,~IEEE}, and
    Yindi Jing,~\IEEEmembership{Member,~IEEE}

   \thanks{
        M.\ Mohammadkarimi, M. Mehrabi, M. Ardakani, and Y. Jing are with the
        Department of Electrical and Computer Engineering, University of Alberta, Edmonton, AB, Canada.
        (e-mail: \texttt{\{mostafa.mohammadkarimi, mehrtash, ardakani,yindi\}@ualberta.ca}).
	}
	\thanks{
	This research was supported by the Huawei Innovation Research Program (HIRP).
	}
	\thanks{
	}	
	
}

\markboth{This paper published in IEEE Transactions on Wireless Communications (DOI: 10.1109/TWC.2019.2924220).}
{Shell \MakeLowercase{\textit{et al.}}: Bare Demo of IEEEtran.cls for IEEE Journals}

\maketitle
	
	\setcounter{page}{1}
	
\begin{abstract}
In this paper,
a \ac{dl}-based sphere decoding algorithm is proposed, where the radius of the decoding hypersphere is learned by a \ac{dnn}.
The performance achieved by the proposed algorithm is very close to the optimal \ac{mld} over a wide range of signal-to-noise ratios (SNRs), while  the computational complexity, compared to existing sphere decoding variants, is significantly reduced.
 This improvement is
attributed to \ac{dnn}'s ability of intelligently learning the radius of the hypersphere used in decoding.
The expected complexity of the proposed \ac{dl}-based algorithm is analytically derived and compared with existing ones. It is shown that the number of
lattice points inside the decoding hypersphere drastically reduces in the \ac{dl}-based algorithm in both the average and worst-case senses.
The effectiveness of the proposed algorithm is shown through simulation for high-dimensional \ac{mimo} systems, using high-order modulations.
\end{abstract}


\acresetall		
	
\section{Introduction}\label{sec:intro}
\IEEEPARstart{T} {he} problem of optimum \ac{mld} in spatial multiplexing \ac{mimo} systems leads to an integer
\ac{ls} problem, which is equivalent to finding the closest lattice point to a given point \cite{jalden2012sphere, mietzner2009multiple,hassibi2005sphere}.
The integer \ac{ls} problem is much more challenging compared to the conventional \ac{ls} problem, where the unknown vector is an arbitrary complex vector, and the solution is easily obtained through pseudo inverse. This is because of the
discrete search space of the integer LS problem which makes it NP hard in both the worst-case sense and the average sense \cite{hassibi2005sphere}.

Various suboptimal solutions, such as zero-forcing (ZF) receiver, \ac{mmse} receiver, \ac{nl}, \ac{nl} with optimal ordering, have been proposed for the integer \ac{ls} problem in \ac{mimo} systems with reduced computational complexity  \cite{hassibi2000efficient}.
These solutions first solve the unconstrained \ac{ls} problem and then perform simple rounding to obtain a feasible lattice point. While these solutions result in cubic-order complexity, their  performance is significantly worse than the optimal solution.

The idea of sphere decoding for \ac{mimo} detection was introduced in \cite{viterbo1999universal}.
Sphere decoding suggests to confine the search space of the original integer \ac{ls} problem to a hypersphere and implement a branch-and-bound search over a tree to achieve \ac{mld} performance.
It can reduce the number of lattice points to be trialled, thus lower the complexity.
Naturally, choosing an appropriate radius for the decoding hypersphere is crucial for sphere decoding.
If the radius is too small, there might not be any lattice point inside the hypersphere. On the other hand, an overly large radius may result in too many lattice points in the hypersphere, hence increasing the decoding complexity.
For example, the choice of radius based on the Babai estimate guarantees the existence of at least one lattice point inside the hypersphere \cite{zhao2009radius}; however, it may lead to a large number of points within the hypersphere. To achieve the exact \ac{mld} performance with reduced complexity, \ac{spi} was proposed in \cite{hochwald2003achieving,hassibi2005sphere}.

Many variations of sphere decoding with reduced computational complexity have also been proposed in the literature \cite{agrell2002closest,chan2002new,barbero2008fixing,shim2008sphere,gowaikar2007statistical,shim2007radius,chang2013effects,zhao2005sphere,vikalo2004iterative,yang2005new}.
Complexity reduction in sphere decoding through
lattice reduction, geometric and probabilistic tree pruning, and $K$-based lattice selection methods have been addressed in \cite{agrell2002closest,chan2002new,barbero2008fixing,shim2008sphere,gowaikar2007statistical,shim2007radius,chang2013effects,zhao2005sphere,vikalo2004iterative,yang2005new}.
On the other hand, a few studies have addressed the problem of radius selection in sphere decoding \cite{hassibi2005sphere,zhao2009radius}. A method to determine the
radius of the decoding hypersphere was proposed in \cite{hassibi2005sphere}. The proposed algorithm chooses the radiuses based on the noise statistics; however, it ignores the effect of the fading channel matrix.
A modified version of radius selection based on Babai estimate has been developed in \cite{zhao2009radius}. The proposed method can solve the problem of sphere decoding failure due to rounding error in floating-point computations.
To take the advantage of sphere decoding for high-dimensional \ac{mimo} systems with
high-order modulations and other applications, such as multi-user communications, massive \ac{mimo}, and relay communications \cite{wang2017performance,mahdaviani2012raptor,sun2012training}, a promising solution is to develop an intelligent mechanism for radius selection to reduce computational complexity without performance degradation.

Recent studies show that \ac{dl} can provide significant performance improvement in signal processing and communications problems  \cite{ye2018power,nachmani2018deep,farsad2017detection,kim2018deep,wang2017deep,samuel2018learningyu,samuel2018learning,o2017deep,he2018model,
o2017introduction,dorner2018deep,felix2018ofdm,aoudia2018end}.
Specifically, \ac{dl} techniques have been employed to improve certain parts of conventional communication systems, such as decoding, modulation, and more \cite{ye2018power,nachmani2018deep,farsad2017detection}.
These improvements are related to the intrinsic property of a \ac{dnn}, which is a
universal function approximator with superior logarithmic learning ability and convenient optimization capability \cite{haykin1994neural,goodfellow2016deep,gybenko1989approximation}.
Besides, existing signal processing algorithms in communications,
while work well for systems with tractable mathematical models, can become inefficient for complicated and large-scale systems with large amount of imperfections and high nonlinearities. Such scenarios
can be solved through \ac{dl}, which can characterize imperfections and nonlinearities via well-structured approximations \cite{wang2017deep,samuel2018learningyu,samuel2018learning}. Moreover, physical layer communication based on the concept of autoencoder has been investigated in \cite{o2017introduction,dorner2018deep,felix2018ofdm,aoudia2018end,o2017deep,he2018model}.

Specifically, the problem of \ac{mimo} detection through \ac{dl} has been investigated in \cite{samuel2018learningyu,samuel2018learning,o2017deep,he2018model}.
The authors in \cite{samuel2018learningyu,samuel2018learning} proposed the
DetNet architecture for \ac{mimo} detection which can achieve near \ac{mld} performance with lower computational complexity without any  knowledge  regarding  the SNR  value.
The joint design of encoder  and decoder using \ac{dl} autoencoder for \ac{mimo} systems was explored in \cite{o2017deep}.
The authors showed that autoencoder
demonstrates significant potential, and its performance approaches the conventional methods.
 The problem of \ac{mimo} detection in time-varying and spatially correlated fading channels was investigated in \cite{he2018model}.
The authors employed \ac{dl} unfolding to improve the iterative \ac{mimo} detection algorithms.

Motivated by these facts,  a sphere decoding algorithm based on \ac{dl} is proposed in this paper, where the radius of the decoding hypersphere is learned by a \ac{dnn} prior to decoding. The \ac{dnn} maps a sequence of
the fading channel matrix elements and the received signals at its input layer into a sequence of learned radiuses at its output layer. The \ac{dnn} is trained in an off-line procedure for the desired \ac{snr} once and is used for the entire communication phase.

Unlike the \ac{spi} algorithm, the proposed \ac{dl}-based algorithm restricts the number of sequential sphere decoding implementations to a maximum predefined value.
Moreover, since  the decoding radiuses are intelligently learnt by a \ac{dnn}, the number of lattice points that lies inside the hypersphere remarkably decreases, which
significantly reduces the computational complexity. On the other hand, the probability of failing to find a solution is close to zero.
To the best of our knowledge, this is the first work in the literature that proposes a mechanism for radius selection dependent on both the fading channel matrix and the noise statistics.

The remaining of this paper is organized as follows. Section
\ref{sec:sm} briefly introduces the basics of \ac{dl} and \ac{dnn}. Section \ref{sec:sm2} presents the system model. Section \ref{tiugkhgkjhkj} describes the
proposed \ac{dl}-based sphere decoding algorithm. In Section \ref{yyiyyjvchghjvjho008087}, an analytical expression for the expected complexity of the proposed algorithm is derived.
Simulation results
are provided in Section \ref{8868548641n43v453}, and conclusions are drawn in
Section \ref{5629479274-2852}.

\subsection{Notations}
Throughout the paper, $(\cdot)^*$ is used for
the complex conjugate, $(\cdot)^T$ is used for transpose, $(\cdot)^{\rm{H}}$ is used for Hermitian, $| \cdot |$ represents the absolute value operator\footnote{For the sets, it represents the cardinality.} $\lfloor \cdot \rceil$ is the operation that rounds a number to its closest integer, $\emptyset$ denotes the empty set, $\E{\cdot}$ is the statistical expectation, $\hat{{x}}$ is an estimate of $x$,
the symbol $\bf{I}$ denotes the
identity matrix, and
the Frobenius  norm of vector $\bf{a}$ is denoted by ${\|}{\bf{a}}{\|}$.

 The inverse of matrix $\bf{A}$ is denoted by
${\bf{A}}^{-1}$. $\Re\{\cdot\}$ and $\Im\{{\cdot}\}$ denote real and imaginary operands, respectively. The gradient operator is denoted by $\nabla$.
The $m$-dimensional complex, real, and complex integer spaces
are denoted by $\mathbb{C}^m$, $\mathbb{R}^m$, and $\mathbb{CZ}^m$, respectively.
Finally, the circularly symmetric complex Gaussian distribution with mean vector $\bm{\mu}$ and covariance matrix $\bf{\Sigma}$ is denoted by $\mathcal{CN}\big{(}\bm{\mu},\bf{\Sigma}\big{)}$, and $\mathcal{U} (a,b)$ represents the continues uniform distribution over the interval $[a,b]$.

\begin{figure}[t]\label{figcc}
\centering
    \includegraphics[width=7cm]{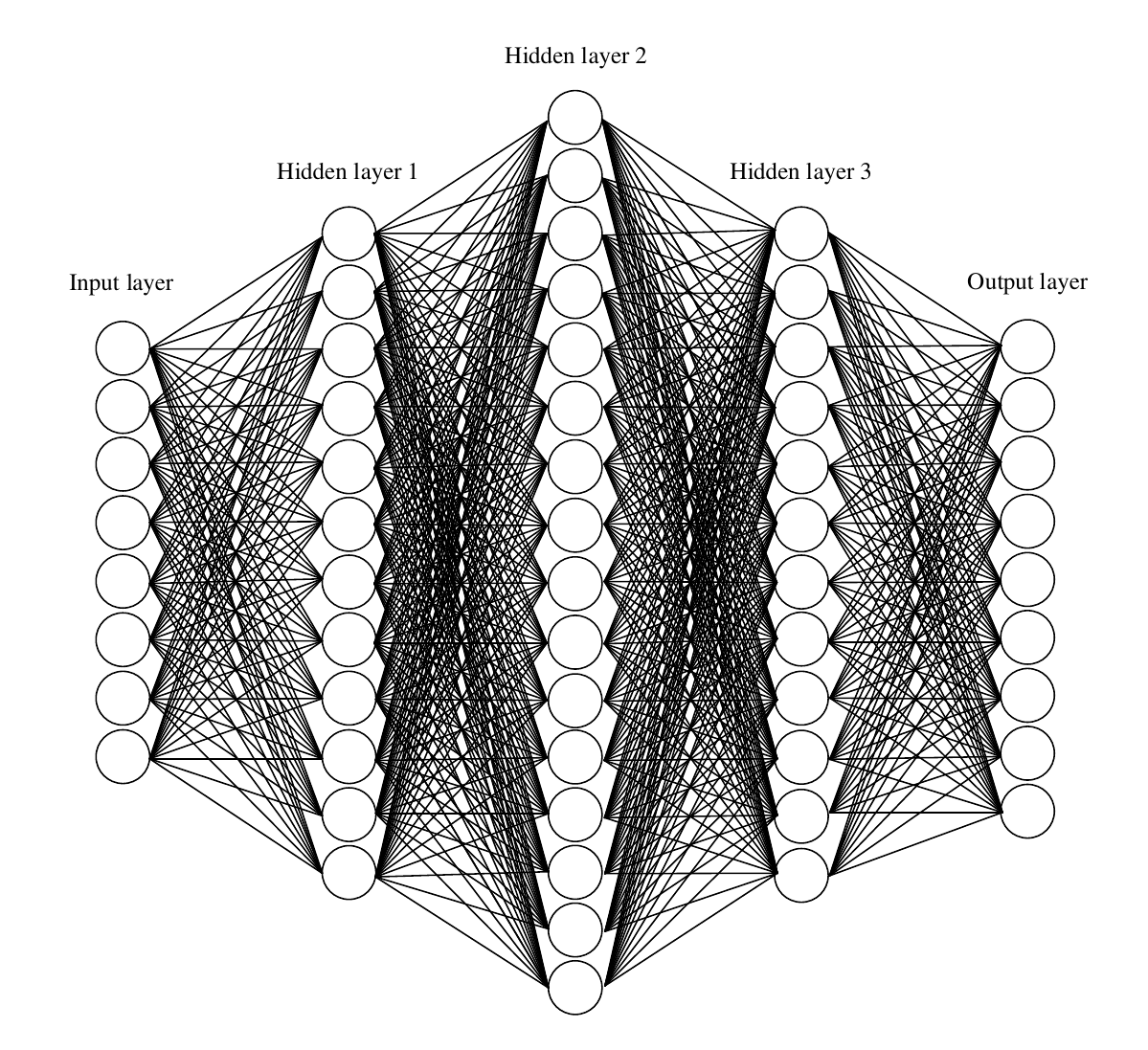}
    \caption{A typical \ac{dnn} with three hidden layers.}\label{fig:long_shortyu}
\end{figure}

\section{\ac{dnn} for Deep Learning}\label{sec:sm}
Deep learning is a subset of artificial intelligence and machine learning that uses multi-layered nonlinear processing units for feature extraction and
transformation. On the contrary to the conventional machine learning techniques, the performance of the DL techniques
significantly improve as the number of training data increases.
Most of the modern \ac{dl} techniques
have been developed based on artificial neural network and are referred to as \ac{dnn}.

A \ac{dnn} is a fully connected feedforward \ac{nn} composed of several hidden layers
and the neurons between the input and output layers. It is distinguished from the conventional \ac{nn} by its depth, i.e., the
number of hidden layers and the number of neurons.
A larger number of hidden layers and neurons enables a \ac{dnn} to extract more meaningful features and
patterns from the data. From a mathematical point of view,
a \ac{nn} is a "universal approximator",
because it can learn to approximate any function ${\bf{z}}=\Upsilon(\bf{x})$ mapping the input vector ${\bf{x}} \in \mathbb{R}^m$ to the output vector ${\bf{z}} \in \mathbb{R}^n$. By employing a cascade of $L$ nonlinear transformations on the input $\bf{x}$, a \ac{nn} approximates
$\bf{z}$ as
\begin{align}
{\bf{z}} \thickapprox T_{(L)}\Big{(}T_{(L-1)}\big{(}\cdots T_1({\bf{x}};\V{\theta}_1);\V{\theta}_{L-1}\big{)};\V{\theta}_{L}\Big{)},
\end{align}
with
\begin{align}\label{uiououoiui98081}
T_{(\ell)}\big{(}{\bf{x}};{\V{\theta}}_\ell\big{)} \triangleq A_\ell\big{(}{\bf{W}}_\ell{\bf{x}}+{\bf{b}}_\ell\big{)},\,\,\,\,\,\ \ell=1,\cdots, L,
\end{align}
where ${\V{\theta}}_\ell \triangleq \big{(} {\bf{W}}_\ell \ {\bf{b}}_\ell \big{)}$  denotes the set of parameters, $ {\bf{W}}_\ell \in \mathbb{R}^{n_\ell\times n_{\ell-1}}$ (where $n_{0}=m$, $n_{L}=n$) and ${\bf{b}}_\ell \in \mathbb{R}^{n_\ell}$  represents the weights and biases, and $A_\ell$ is the activation function of the $l$th layer.
The activation function is applied at each neuron to produce non-linearity. The weights and biases are usually learned through a training set with known desired outputs \cite{goodfellow2016deep}. Fig. \ref{fig:long_shortyu} shows a typical \ac{dnn} with three hidden layers.

\section{System Model}\label{sec:sm2}

We consider a spatial multiplexing \ac{mimo} system with $m$
transmit and $n$ receive antennas. The
vector of received basedband symbols, ${\bf{y}} \in \mathbb{C}^{n\times 1}$, in block-fading channels is modeled as
\begin{align}\label{eq:linear}
\bf{y}=\bf{H}\bf{s}+\bf{w},
\end{align}
where ${\bf{s}}=[s_1,s_2,\cdots,s_m]^T \subset \mathbb{CZ}^m$ denotes the vector of transmitted complex
symbols drawn from an arbitrary constellation $\mathbb{D}$, ${\bf{H}} \in \mathbb{C}^{n\times m}$ is the channel matrix, and ${\bf{w}} \in \mathbb{C}^{n\times 1}$ is the zero-mean \ac{awgn} with covariance matrix ${\bf{\Sigma}}_{\rv{w}}=\sigma_{\rm{w}}^2\bf{I}$.  The channel from transmit antenna $j$ to receive antenna $i$ is denoted by $h_{ij}$.

The vector $\bf{s}$ spans the "rectangular" $m$-dimensional complex integer lattice $\mathbb{D}^m \subset  \mathbb{CZ}^m$, and the $n$-dimensional vector $\bf{H}\bf{s}$ spans a "skewed" lattice for any given lattice-generating matrix $\bf{H}$.
With the assumption that $\bf{H}$ is perfectly estimated at the receiver, \ac{mld} of the vector $\bf{s}$ in \eqref{eq:linear} given the observation vector $\bf{y}$,  leads to the following integer \ac{ls} problem:
\begin{align}
\underset{{\bf{s}} \in \mathbb{D}^m \subset  \mathbb{CZ}^m} {\min} \,\,\,\ \big{\Vert}{\bf{y}}-{\bf{H}}{\bf{s}}\big{\Vert}^2. \label{eq:uuion2}
\end{align}
As seen, the integer \ac{ls} problem in \eqref{eq:uuion2} is equivalent to finding the closest point in the skewed lattice $\bf{H}\bf{s}$ to the vector $\bf{y}$ in the Euclidean sense. For large values of $m$ and high-order modulation, exhaustive search is computationally unaffordable.

{Sphere decoding can speed up the process of finding the optimal solution
by searching only the points of the skewed lattice that lie within a hypersphere of
radius $d$ centered at the vector $\bf{y}$. This can be mathematically expressed as}
\begin{align}\label{uui6773mn}
    \displaystyle{\min_{\substack{{\bf{s}} \in \mathbb{D}^m \subset  \mathbb{CZ}^m \\ {\Vert{\bf{y}}-{\bf{H}}{\bf{s}}\Vert^2} \leq d^2 }}}
  \big{\|}{\bf{y}}-{\bf{H}}{\bf{s}}\big{\|}^2.
\end{align}
{It is obvious that the closest lattice point inside the hypersphere is also the closest point for the whole lattice.
The main problem in sphere decoding is how to choose
$d$ to avoid a large number of lattice points inside the hypersphere and at the same time guarantee the existence of a lattice point inside the hypersphere for any vector $\bf{y}$.}

To achieve \ac{mld} error performance, \ac{spi} is required since
for any hypersphere radius $r_i$, there is always a non-zero probability that
this hypersphere does not contain any lattice point. When no lattice point is available, the search radius needs to be increased from $r_i$ to $r_{i+1}$, and the search is conducted again. This procedure continues until the optimal solution is obtained.
While \ac{spi} substantially improve on the complexity of \ac{mld}
from  an  implementation  standpoint, the average and worst-case complexity can still be huge when there are no lattice points in the hypersphere with radius $r_i$, but many in the hypersphere with radius $r_{i+1}$. Hence, the choice of $r_i$'s is critical.

\section{\ac{dl}-based Sphere Decoding}\label{tiugkhgkjhkj}
The main idea behind our proposed \ac{dl}-based sphere decoding algorithm is to implement \ac{spi} for a small number of intelligently learned radiuses. That is, $r_i$'s are learned and chosen intelligently by a DNN.
\ac{dl}-based sphere decoding makes it possible to choose the decoding radiuses based on the noise statistics and the structure of $\bf{H}$. This significantly increases the probability of successful \ac{mld} with searching over only a small number of lattice points.

In the proposed \ac{dl}-based sphere decoding,
 the Euclidean distance of the $q$ closest lattice points to vector $\bf{y}$ in the skewed lattice space is reconstructed via a \ac{dnn} (as the \ac{dnn} output) prior to sequential sphere decoding implementations.
Then, these $q$ learned Euclidean distances are used as radiuses of the hyperspheres in sphere decoding implementations.
 The value of $q$ is chosen small due to computational complexity consideration.
Ideally, if the distances are produced with no error, $q=1$ is sufficient for the optimal decoding with the lowest complexity,
since the radius is the distance of $\bf{y}$  to the optimal \ac{mld} solution. This radius for sphere decoding guarantees the existence of a point inside the hypersphere  and actually only the optimal point is inside the hypersphere.
 However, since a \ac{dnn} is an approximator, there is the possibility that no points lies within the hypersphere with the learned radius. Thus, instead of learning the closest distance only,
$q$ closest Euclidean distances are learnt by the \ac{dnn} to increase the probability of finding the optimal lattice point.
Since for any finite value of $q$, still there is the possibility that no lattice point lies within the hypersphere with the largest learned radius, a suboptimal detector, such as \ac{mmse} with rounding or \ac{nl} with optimal ordering
is employed in order to avoid failure in decoding.

Let us define the  Euclidean distance between $\bf{y}$ and the $i$th lattice point in the skewed lattice, i.e., ${\bf{H}}{\bf{s}}_i$, as
\begin{align}\label{gtrjhgljlgjttww}
{r}_i\triangleq{\big{\Vert}{\bf{y}}-{\bf{H}}{\bf{s}}_i\big{\Vert}},\,\,\,\,\,\ i= 1,2,\cdots,|\mathbb{D}|^{m},
\end{align}
where $|\mathbb{D}|$ is the cardinality of the constellation $\mathbb{D}$. Further,
by ordering $r_i$ as follows,
\begin{align}
{r}_{i_1} < {r}_{i_2} < \cdots < {r}_{i_q} < {r}_{i_{q+1}}  < \cdots < {r}_{i_{|\mathbb{D}|^m}},
\end{align}
the  desired $q \times 1$ radius vector ${\bf{r}}$ is given as
\begin{align}\label{yyyerw234cvx}
{\bf{r}} \triangleq  [{r}_{i_1} \ {r}_{i_2} \ \cdots  \ {r}_{i_q} ]^T.
\end{align}
In the proposed \ac{dl}-based sphere decoding algorithm,
the \ac{dnn}, $\Phi(\bf{x};\V{\theta})$, reconstructs the radius vector $\bf{r}$ at its output layer as
\begin{align}
{\hat{\bf{r}}}=\Phi(\bf{x};\V{\theta}),
\end{align}
where
\begin{align}\label{eq22223}
{\bf{x}} \triangleq \Big{[}\bar{\bf{y}}  \  \tilde{\bf{y}} \ {\bar{h}}_{11} \ {\tilde{h}}_{11}  \cdots \ {\bar{h}}_{nm} \ {\tilde{h}}_{nm} \Big{]}^T,
\end{align}
$\bar{\bf{y}}=\Re{\{\bf{y}}^T\}$, $\tilde{\bf{y}}=\Im{\{\bf{y}}^T\}$, $h_{uv}\triangleq \bar{h}_{uv}+i\tilde{h}_{uv}$, and $\V{\theta} \triangleq [{\theta}_1,{\theta}_2, \cdots, {\theta}_K]^T$. The vector $\bf{x}$ represents the input vector of the \ac{dnn}, and $\V{\theta}$
is the vector of all parameters of the \ac{dnn}.

The proposed \ac{dl}-based sphere decoding is composed of an off-line training phase, where the parameters of the \ac{dnn} is obtained by employing training examples,
and a decoding phase where the transmit vector is decoded through sphere decoding or a suboptimal detector.
In the following subsection, these two phases are explained in details.
\begin{algorithm}[t]\label{mpl1l2}
    \caption{\ac{dl}-based sphere decoding algorithm}
    \begin{algorithmic}[1]
    \Statex \textbf{Input:} $\bf{y}$, $\bf{H}$, $\Phi(\cdot,\V{\theta})$, $q$
      \Statex \textbf{Output:} $\hat{\bf{s}}$
      \State Stack $\bf{y}$ and $\bf{H}$ as in \eqref{eq22223} to obtain $\bf{x}$;
      \State Obtain the $q$ radiuses through the trained \ac{dnn} as ${\hat{\bf{r}}}=\Phi({\bf{x}},\V{\theta})=[{\hat{r}}_{i_1} \ {\hat{r}}_{i_2} \ \cdots  \ {\hat{r}}_{i_q} ]^T$;
      \State $c=1$;
      \State Implement sphere decoding for radius $\hat{r}_{i_c}$;
      \State \textbf{if} $D_{\rm{sp}}({\bf{y}},\hat{r}_{i_c})\neq {\text{null}}$
       \State \ \ \ $\hat{\bf{s}}=D_{\rm{sp}}({\bf{y}},\hat{r}_{i_c})$;
       \State \textbf{else}
       \Statex \ \ \ \ \ \ \ \ \ \ \ \textbf{if} $c<q$
       \State \ \ \ \ \ \ \ \ \ \ \ \ \ \ \ $c=c+1$ and go to 4;
       \State \ \ \ \ \ \ \ \ \ \ \ \textbf{else}
        \State \ \ \ \ \ \ \ \ \ \ \ \ \ \ \ $\hat{\bf{s}}=D_{\rm{sb}}(\bf{y})$;
        \State \ \ \ \ \ \ \ \ \ \ \ \textbf{end}
        \State \textbf{end}
\end{algorithmic} \label{euclidendwhilemmmmmeqqqq}
  \end{algorithm}
  \vspace{-2em}
\subsection{Training Phase}
A three layers \ac{dnn} with one hidden layer is considered for a $10 \times 10$ spatial multiplexing \ac{mimo} system using $16$-QAM and $64$-QAM in the training phase,\footnote{Based on the \ac{mimo} configuration and modulation type, different \ac{dnn} architecture can be selected.}  where the numbers of neurons in each layers are $220$, $128$, and $q$, respectively.
Clipped rectified linear unit with the following mathematical operation is used as the activation function in the hidden layers:
\begin{equation}
  f(u)=\begin{cases}
    0, & u<0 .\\
    u, & 0\leq u<1 \\
    1, & u\geq 1
  \end{cases}.
\end{equation}
Table \ref{uuuuionm098} summarizes the architecture of the employed \ac{dnn} for the $10 \times 10$ spatial multiplexing \ac{mimo} system in this paper.
	\begin{table}[]
		\vspace{.1cm}
		\centering
		\caption{Layout of the employed \ac{dnn}.}
		\begin{tabular}{l*{6}{c}r}
			Layer              & Output Dimension &  Parameters  \\
			\hline
			Input & 220 & 0  \\
			Dense + CRelu & 128 & 28,288  \\
			Dense & $q$ & $129 \times q$   \\
			\hline
		\end{tabular}\label{uuuuionm098}
	\end{table}

It should be mentioned that an \ac{snr} dependent \ac{dnn}, in which the structure of the \ac{dnn} is designed to be adaptive to the \ac{snr} value,
can also be employed to further reduce computational complexity. For the sake of simplicity, a fixed \ac{dnn} is used
for all \ac{snr} values  in this paper. However, the network is independently trained for each \ac{snr} value.

In the training phase, the designed \ac{dnn} is trained with independent input vectors, given as
\begin{align}\label{rrtcmc1212}
{\bf{x}}^{(i)} \triangleq   \Big{[}\bar{\bf{y}}^{(i)}  \  \tilde{\bf{y}}^{(i)} \ {\bar{h}}_{11}^{(i)} \ {\tilde{h}}_{11}^{(i)}  \cdots \ {\bar{h}}_{nm}^{(i)} \ {\tilde{h}}_{nm}^{(i)} \Big{]}^T
\end{align}
for $i=1,2,\cdots,N$
to obtain
the parameter vector $\V{\theta}$ of the \ac{dnn} by minimizing the following \ac{mse} loss function \cite{o2017deep}:
\begin{align}\label{yyy:008}
Loss(\V{\theta})\triangleq \frac{1}{N}\sum_{i=1}^{N} \Big{\|}{\bf{r}}^{(i)}-\Phi({\bf{x}}^{(i)};\V{\theta})\Big{\|}^2,
\end{align}
where ${\bf{r}}^{(i)}$ is the desired radius vector when ${\bf{x}}^{(i)}$ is used as input vector.
To achieve faster convergence and decrease computational complexity, an approximation of the \ac{mse} loss function in \eqref{yyy:008} is computed for random mini-batches of training examples at each iteration $t$ as
\begin{align}\label{yyy:008hjg}
f_t(\bm{\theta})\triangleq \frac{1}{M}\sum_{i=1}^M \Big{\|}{\bf{r}}^{(M(t-1)+i)}-\Phi({\bf{x}}^{(M(t-1)+i)};\bm{\theta})\Big{\|}^2,
\end{align}
where  $M$ is the mini-batch size, and $B=N/M$ is the number of batches.
The training data is randomly shuffled before every epoch.\footnote{Each epoch is one forward pass and one backward pass of all the training examples.}
 By choosing $M$ to be considerably small compared to $N$, the complexity of the gradient computation for one epoch, i.e., $\nabla_{\bm{\theta}} f_t(\bm{\theta}_{t-1})$, $t=1,2,\cdots,B$, remarkably decreases when compared to $\nabla_{\bm{\theta}} Loss(\V{\theta})$, while the variance of the parameter update still decreases.

During the training phase, for each \ac{snr} value, elements of the transmitted vector ${\bf{s}}^{(i)}$, elements of the fading channel matrix ${\bf{H}}^{(i)}$, and elements of the noise vector ${\bf{w}}^{(i)}$,  $i=1, \cdots, N$, are independently and uniformly drawn from $\mathbb{D}$, $f_{\rm{h}}({h})$, and ${\cal{CN}}\big{(}{0},\sigma_{\rm{w}}^2\big{)}$, respectively, where $f_{\rm{h}}({h})$ denotes the distribution of the fading channel.
Then, the real and imaginary parts of the observation vectors during training, i.e., ${\bf{y}}^{(i)}={\bf{H}}^{(i)}{\bf{s}}^{(i)}+{\bf{w}}^{(i)}$, along with ${\bf{H}}^{(i)}$ are stacked as in \eqref{rrtcmc1212} and fed to the \ac{dnn} to minimize the \ac{mse} loss function in \eqref{yyy:008hjg}. For each input training vector  ${\bf{x}}^{(i)}$, the
corresponding desired radius vector  ${\bf{r}}^{(i)}$ is obtained by employing \ac{spi} with a set of heuristic radiuses.
Finally, the parameter vector of the \ac{dnn} is updated according to the input-output vector pairs $({\bf{x}}^{(i)},{\bf{r}}^{(i)})$ by employing the adaptive moment estimation
stochastic optimization algorithm, which is also referred to as Adam algorithm \cite{kingma2014adam}.

Since the \ac{dnn} in our algorithm is used as a function approximator, the \ac{dnn} does not need to see all possible codewords in the training phase. The \ac{dnn} only approximates the region in which the optimal solution exists. Hence, the number of training samples, and thus, the computational complexity of the training can be linearly scaled up by the cardinality of constellation.

\subsection{Decoding Procedure}
In the decoding phase, first, the received vector $\bf{y}$ and fading channel matrix $\bf{H}$ are fed to the trained \ac{dnn} in the form of \eqref{eq22223} to produce the radius vector $\hat{\bf{r}}\triangleq [{\hat{r}}_{i_1} \ {\hat{r}}_{i_2} \ \cdots  \ {\hat{r}}_{i_q} ]^T$; then, the transmitted signal vector is decoded by Algorithm \ref{euclidendwhilemmmmmeqqqq},
where sphere decoding is conducted recursively with the learned radiuses by the \ac{dnn}, followed by a suboptimal detection if the sphere decoding fails to find the solution. Sphere decoding implementation with decoding radius $\hat{r}_{i_c}$ fails to find a solution when
\begin{align}
\setcounter{equation}{16}
\Set{C}({\bf y}, \hat{r}_{i_c}) \triangleq
\Big{\{} {\bf s} \in \mathbb{D}^m \big{|}  \big{\|}{\bf y} - {\bf H} {\bf s} \big{\|}^2 \le \hat{r}_{i_c}^2 \Big{\}}
=\emptyset
\end{align}
that is, there is no lattice point inside the hypersphere with radius $\hat{r}_{i_c}$. To help the presentation, define
  \begin{equation}
    D_{\rm{sp}}({\bf{y}},\hat{r}_{i_c}) \triangleq
    \begin{cases}
    \displaystyle{\min_{\substack{{\bf{s}} \in \mathbb{D}^m \subset  \mathbb{CZ}^m \\ {\Vert{\bf{y}}-{\bf{H}}{\bf{s}}\Vert^2} \leq \hat{r}_{i_c}^2 }}
  \big{\Vert}{\bf{y}}-{\bf{H}}{\bf{s}}\big{\Vert}^2}, & \text{if}\ \Set{C}({\bf y}, \hat{r}_{i_c}) \ne \emptyset  \\
     \hspace{2em} {\text{null}}, & \text{if}\ \Set{C}({\bf y}, \hat{r}_{i_c}) = \emptyset.
    \end{cases}
  \end{equation}
On the other hand, if no point is found by the $q$ rounds of sphere decoding, \ac{mmse} is  employed as the suboptimal detector, in which the solution is obtained as
\begin{align}\label{88fjbbk785rrrrttyy}
 D_{\rm{sb}}({\bf{y}})=\Big{\lfloor}({\bf{H}}^{\rm{H}}{\bf{H}}+\bar{\gamma}^{-1}{\bf{I}})^{-1}{\bf{H}}^{\rm{H}}\bf{y}\Big{\rceil},
\end{align}
where $\bar{\gamma}$ is the average \ac{snr}.
Simulation result show that due to the intelligent production of the radiuses via a DNN,
the probability of decoding through suboptimal detector is very close to zero.

\subsection{Intuition behind \ac{dl}-based Sphere Decoding}
Since the complexity of sphere decoding algorithm is data dependent (depends on ${\bf{y}}$ and ${\bf{H}}$), data dependent hypersphere radius selection can lead to lower computational complexity \cite{giannakis2007space,hassibi2005sphere}.
The proposed \ac{dl}-based sphere decoding algorithm selects the hypersphere radiuses dependent on ${\bf{y}}$ and ${\bf{H}}$ to reduce computational complexity.

The \ac{nn} in the proposed \ac{dl}-based sphere decoding behaves as a function approximator.
In the mathematical theory of artificial neural networks, the universal approximation theorem
states \cite{haykin1994neural} that a feed-forward network with a single hidden layer containing a finite number of neurons
can provide an arbitrarily close approximation to a continuous function $f(\bf{x})$, on compact subsets of $\mathbb{R}^n$, under mild assumptions on the
activation function. A formal description of this theorem is provided below.

\begin{theorem}(Universal Approximation Theorem):

Let $\varphi(\cdot)$ $\mathbb{R} \rightarrow \mathbb{R}$ be a nonconstant, bounded and continuous function.
Then, given any $\epsilon > 0$  and any function $f: \mathbb{I}_m \rightarrow \mathbb{R}$, where $\mathbb{I}_m$  is a compact subset of $\mathbb{R}^m$,
there exist an integer $N$, real constants $v_i$, $b_i \in \mathbb{R}$, and real vectors ${\bf{w}}_i \in \mathbb{R}^m$ for $i = 1 , \cdots, N$, such that
\begin{align}
| F( {\bf{x}} ) - f ( {\bf{x}} ) | < \epsilon,
\end{align}
where
\begin{align}
  F( {\bf{x}} ) =
  \sum_{i=1}^{N} v_i \varphi \left( {\bf{w}}_i^T {\bf{x}} + b_i\right).
\end{align}
$F( {\bf{x}} ) $ is an approximate realization of the function $f$. This result holds even if the function has many outputs.
A visual proof that \ac{nn} can approximates any continues function is provided in \cite{gybenko1989approximation,WinNT}.
\end{theorem}

Based on this theorem, the employed \ac{nn} in our algorithm approximates the function ${\bf{r}}={\bf{g}}({\bf{y}},{\bf{H}}), \mathbb{R}^{2n(m+1)}\rightarrow \mathbb{R}^{q}$, where ${\bf{g}}({\bf{y}},{\bf{H}})\hspace{-0.2em} \triangleq \hspace{-0.2em} [g_1({\bf{x}},{\bf{H}})\ g_2({\bf{x}},{\bf{H}})\cdots\ g_q({\bf{x}},{\bf{H}})]^T$ and $g_i({\bf{x}},{\bf{H}})$ is the
distance of the $i$ closest lattice points to vector $\bf{y}$, i.e.,
\begin{align}
g_i({\bf{y}},{\bf{H}})={\big{\|}{\bf{y}}-{\bf{H}}\hat{\bf{s}}_i\big{\|}^2}, \,\,\,\,\,\,\,\,\,\, i=1,2,\cdots,q,
\end{align}
where $g_1({\bf{y}},{\bf{H}})\leq  g_2({\bf{y}},{\bf{H}}) \leq \cdots \leq g_q({\bf{y}},{\bf{H}})$ and
\begin{align}\label{uui6773mn}
\hat{\bf{s}}_i=\displaystyle{\min_{\substack{{\bf{s}} \in \mathbb{D}^m \subset  \mathbb{CZ}^m \\ {\bf{s}} \not\in  \{\hat{\bf{s}}_1, \cdots, \hat{\bf{s}}_{i-1}\} }}}
  \big{\|}{\bf{y}}-{\bf{H}}{\bf{s}}\big{\|}^2.
\end{align}
 The reason that non-linear function $\bf{g}$ is learned is
to gradually and in a controlled manner increase the hypersphere radius to avoid too many lattices inside the search hyperspheres.
The learned \ac{nn} makes data dependent radius selection, and thus,
reduces the number of lattice points that fall in the hyperspheres.
It is worth noting that the complexity of sphere decoding is proportional to the number of lattice points that lies inside the hypersphere \cite{hassibi2005sphere}.

\section{Expected Complexity of the \ac{dl}-based Sphere Decoding}\label{yyiyyjvchghjvjho008087}
In this section, the expected complexity of the proposed \ac{dl}-based sphere decoding algorithm in the decoding pahse is analytically derived. Since the \ac{dnn} is trained once and is used for the entire decoding phase, the expected complexity of the training phase is not considered.

\begin{customlemma}{1}\label{one}
The expected complexity of the proposed \ac{dl}-based sphere decoding algorithm is obtained as
\begin{align}\label{hhhbbccxzzrr56577}
& C_{\rm{DL}}(m,\sigma^2)=\sum_{c=1}^{q}\sum_{k=1}^{m}\sum_{v=0}^{\infty}F_{\rm{sp}}(k)\Psi_{2k}(v) \\ \nonumber
&\times \E{\gamma \Big{(}\frac{\hat{r}_{i_c}^2}{\sigma_{\rm{w}}^2+v},n-m+k\Big{)}\bigg{(}\gamma\Big{(}\frac{\hat{r}_{i_c}^2}{\sigma_{\rm{w}}^2},n\Big{)}-\gamma\Big{(}\frac{\hat{r}_{i_{c-1}}^2}{\sigma_{\rm{w}}^2},n\Big{)}\bigg{)}} \\ \nonumber
&+\bigg{(}1-\mathbb{E}\bigg{\{} \gamma\Big{(}\frac{\hat{r}_{i_q}^2}{\sigma_{\rm{w}}^2},n\Big{)}\bigg{\}}\bigg{)}F_{\rm{sb}}+F_{\rm{dn}},
\end{align}
where $\hat{r}_{i_0}=0$, $\gamma(\cdot,\cdot)$ is the lower incomplete gamma function, $\Psi_{2k}(v)$ is the number of ways that $v$ can be represented as the sum of $2k$ squared integers,
\begin{align}\label{5557880785qasx}
F_{\rm{sb}}=m^3+\frac{5m^2}{2}+nm^2+3mn-\frac{m}{2},
\end{align}
\begin{align}\label{ghssdkhckshdkh3}
F_{\rm{dn}}=\sum_{i=0}^{L-1}2n_{i+1}n_{i},
\end{align}
and $F_{\rm{sp}}(k)$ is the number of elementary operations including complex additions, subtractions, and multiplications per visited point in complex dimension $k$ in sphere decoding.
\end{customlemma}

\begin{proof}\nonumber
See the appendix.
\end{proof}

As seen from the proof in the appendix, $F_{{\rm{sb}}}$ and $F_{\rm{dn}}$ represents the number of elementary operations employed by the \ac{mmse} suboptimal detector in \eqref{88fjbbk785rrrrttyy} and \ac{dnn}, respectively.
Also, the term $\gamma {(}\sfrac{\hat{r}_{{i_c}}^2}{\sigma_{\rm{w}}^2},n{)}$ in \eqref{hhhbbccxzzrr56577} is the probability of finding at least a lattice point inside the hypersphere with the learned radius  $\hat{r}_{i_c}$, which is written as
\begin{align}\label{tgdjgadgagdjh67661w19721}
\hat{p}_{i_c}\triangleq  \gamma\Big{(}\frac{\hat{r}_{i_c}^2}{\sigma_{\rm{w}}^2},n\Big{)}= \int_{0}^{\frac{\hat{r}_{i_c}^2}{\sigma_{\rm{w}}^2}} \frac{t^{n-1}}{\Gamma{(n)}}\exp(-t)dt,
\end{align}
where $\hat{p}_{i_0}=0$.

\begin{figure*}
\begin{align}\nonumber
C_{\rm{DL}}(m,\sigma^2)=&\lim_{U\to\infty} \frac{1}{U} \sum_{u=1}^{U}\sum_{c=1}^{q}\sum_{k=1}^{m}\sum_{v=0}^{\infty}F_{\rm{sp}}(k)\Psi_{2k}(v)
 \times \gamma \Big{(}\frac{\hat{r}_{i_c,u}^2}{\sigma_{\rm{w}}^2+v},n-m+k\Big{)}\bigg{(}\gamma\Big{(}\frac{\hat{r}_{{i_c},u}^2}{\sigma_{\rm{w}}^2},n\Big{)}-\gamma\Big{(}\frac{\hat{r}_{i_{c-1},u}^2}{\sigma_{\rm{w}}^2},n\Big{)}\bigg{)}\\ \label{7788485pp0886}
&+\Big{(}m^3+\frac{5m^2}{2}+nm^2+3mn-\frac{m}{2}\Big{)}\bigg{(}1-\frac{1}{U}\sum_{u=1}^{U}\gamma\Big{(}\frac{\hat{r}_{{i_q},u}^2}{\sigma_{\rm{w}}^2},n\Big{)}\bigg{)}+F_{\rm{dn}}.
\end{align}
\hrulefill
\end{figure*}

By replacing the statistical expectation with sample mean based on Monte Carlo sampling, one can write the expected complexity of the \ac{dl}-based algorithm as in  \eqref{7788485pp0886},
where the subscript $u$ represents the index of sample in importance sampling.

For $M^2$-QAM constellation, $F_{\rm{sp}}(k)=8k+20+4M$, and $\Psi_{2k}(v)$ for 4-QAM, 16-QAM, and 64-QAM is respectively given as \cite{vikalo2005sphere}
\begin{equation}
\setcounter{equation}{29}
    \Psi_{2k}(v)=
    \begin{cases}
      {{2k}\choose{v}}, & \text{if}\ 0\leq v \leq 2k \\
      0 & \text{otherwise},
    \end{cases}
  \end{equation}
\begin{equation}
    \Psi_{2k}(v)=
    \begin{cases}
     \sum\limits_{j=0}^{2k}   \frac{1}{2^{2k}} {{2k}\choose{j}}\Omega_{2k,j}(v), & \text{if}\ v \in \Xi \\
      0 & \text{otherwise},
    \end{cases}
  \end{equation}
and
\begin{equation}
    \Psi_{2k}(v)=
    \begin{cases}
     \sum\limits_{\xi_0,\xi_1,\xi_2,\xi_3} \frac{1}{4^{2k}} \Omega_{2k,\xi_0,\xi_1,\xi_2,\xi_3}(v), & \text{if}\ v \in \Set{Q} \\
      0, & \text{otherwise}
    \end{cases}
  \end{equation}
where $\Omega_{2k,j}(v)$ is the coefficient of $\lambda^v$ in the polynomial
\begin{align}\label{769707098281qa}
(1+\lambda+\lambda^4+\lambda^9)^j(1+2\lambda+\lambda^4)^{2k-j},
\end{align}
the set $\Xi$ contains the coefficients of the polynomial in \eqref{769707098281qa} for $k=1, \cdots,m$ and  $j=0, \cdots, 2k$,
$\Omega_{2k,\xi_0,\xi_1,xi_2,xi_3}(v)$ is the coefficient of $\lambda^v$ in the polynomial
\begin{align}\label{tei2iduh2kjhdlh125}
&{{2k}\choose{\xi_0,\xi_1,\xi_2,\xi_3}}\bigg{(}\sum_{e_0=0}^{7}\lambda^{e_0^2}\bigg{)}^{\xi_0} \bigg{(}\lambda+\sum_{e_1=0}^{6} \lambda^{e_1^2}\bigg{)}^{\xi_1} \\ \nonumber
&\times \bigg{(}\lambda+\lambda^4+\sum_{e_2=0}^{5} \lambda^{e_2^2}\bigg{)}^{\xi_2}\bigg{(}-1-\lambda^{16}+\sum_{e_3=0}^{4} 2\lambda^{e_3^2}\bigg{)}^{\xi_3},
\end{align}
where $\xi_0+\xi_1+\xi_2+\xi_3=2k$, ${{2k}\choose{\xi_0,\xi_1,\xi_2,\xi_3}}=\sfrac{(2k)!}{(\xi_0!\xi_1!\xi_2!\xi_3!)}$,
and the set $\Set{Q}$ contains the coefficients of the polynomial in \eqref{tei2iduh2kjhdlh125} for $k=1, \cdots,m$.

\subsection{Asymptotic Complexity Analysis}
As the \ac{snr} approaches $+\infty$, the \ac{awgn} noise can be ignored. In this case, the expected complexity of the proposed \ac{dl}-based sphere decoding algorithm in \eqref{hhhbbccxzzrr56577} is simplified as
\begin{align}
&C_{\rm{DL}}(m,0) \\ \nonumber
& =\Big{(}m^3+\frac{5m^2}{2}+nm^2+3mn-\frac{m}{2}\Big{)}+\sum_{i=0}^{L-1}2n_{i+1}n_{i}.
\end{align}
Since the number of neurons in the hidden layer is linearly scaled up the size of the input layer $2(m(n+1))$, the complexity order
of the proposed algorithm for large \ac{mimo} systems $(n\geq m>>1)$, is $\mathcal{O}(m^2n^2)$.

\section{Simulation Results}\label{8868548641n43v453}
In this section, we evaluate the performance of the proposed \ac{dl}-based sphere decoding algorithm through
several simulation experiments.

\subsection{Simulation Setup}
We consider a $10 \times 10$ spatial multiplexing \ac{mimo} system in Rayleigh block-fading channel where $16$-QAM and $64-$QAM are employed. These configurations result in skewed lattices with $4^{20}$ and $4^{30}$ lattice points, respectively.
The elements of the fading channel matrix are modeled as \ac{iid} zero-mean  circularly symmetric complex Gaussian random variables with unit variance.
The additive white noise is modeled as a circularly symmetric complex-valued Gaussian random variable with zero-mean and variance $\sigma_{\rm{w}}^2$
for each receive antennas. Without loss of generality, the average \ac{snr} in dB is defined as $\gamma \triangleq 10 \log \big{(}\sfrac{m\sigma_{\rm{s}}^2}{\sigma_{\rm{w}}^2}\big{)}$, where $\sigma_{\rm{s}}^2$ is the average signal power, and $m$ is the number of transmit antennas.
The \ac{nn} is implemented
using Deep learning Toolbox of MATLAB 2019a. The learning rate of the Adam optimization
algorithm is set to 0.001, and the parameters of the employed \ac{nn} in the training phase are given in Table \ref{uuuuionm09fffff8}:
	\begin{table}[]
		\vspace{.1cm}
		\centering
		\caption{Training phase parameters.}
		\begin{tabular}{l*{6}{c}r}
			Parameter             & \textbf{16-QAM} &  \textbf{64-QAM}  \\
			\hline
      Number of batches & 90 & 90   \\
      Size of batches & 200 & 200   \\
      Number of epoches & 20 & 23 \\
      Number of iterations & 1800 & 2070 \\
			\hline
		\end{tabular}\label{uuuuionm09fffff8}
	\end{table}

\begin{figure}[t!]
\centering
\subfloat[EPDF of $\hat{r}_1$]{\includegraphics[height=2.4in]{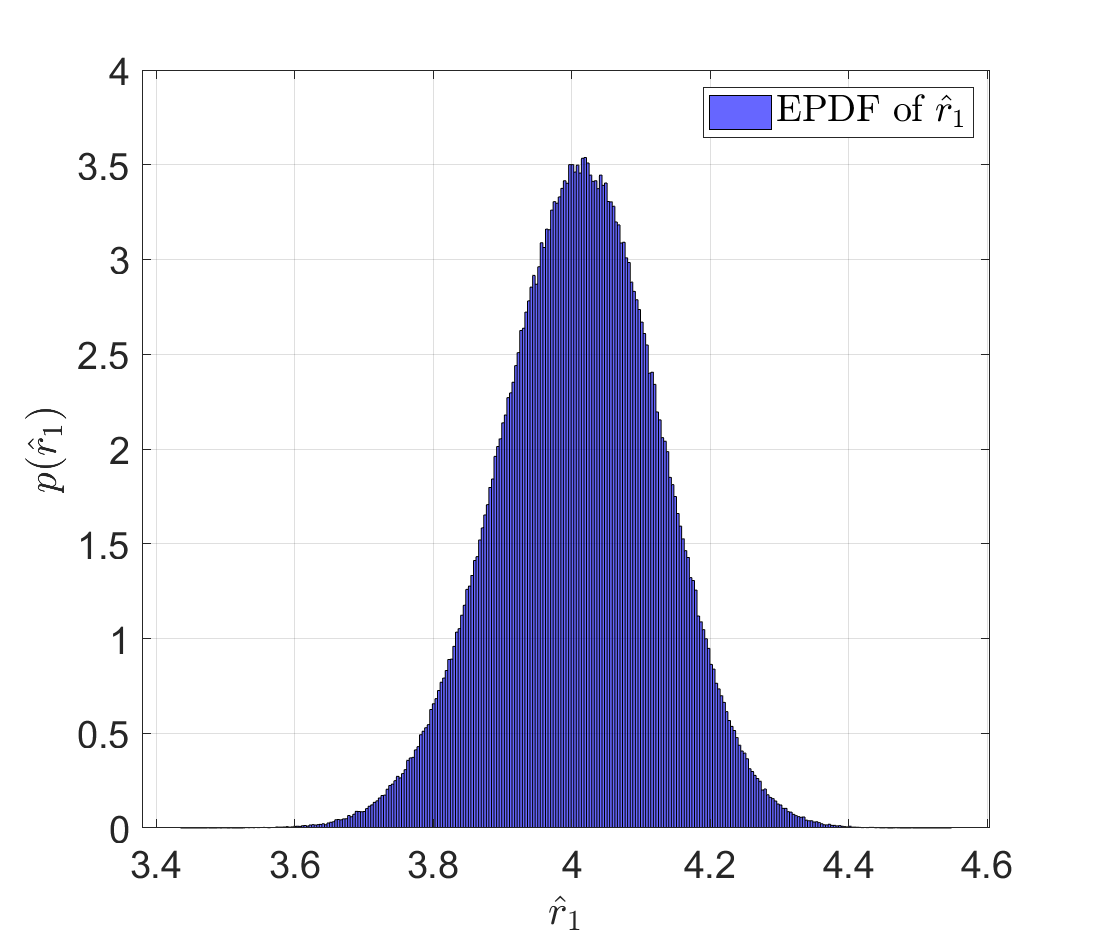}}
\\
\subfloat[EPDF of $\hat{r}_2$]{\includegraphics[height=2.4in]{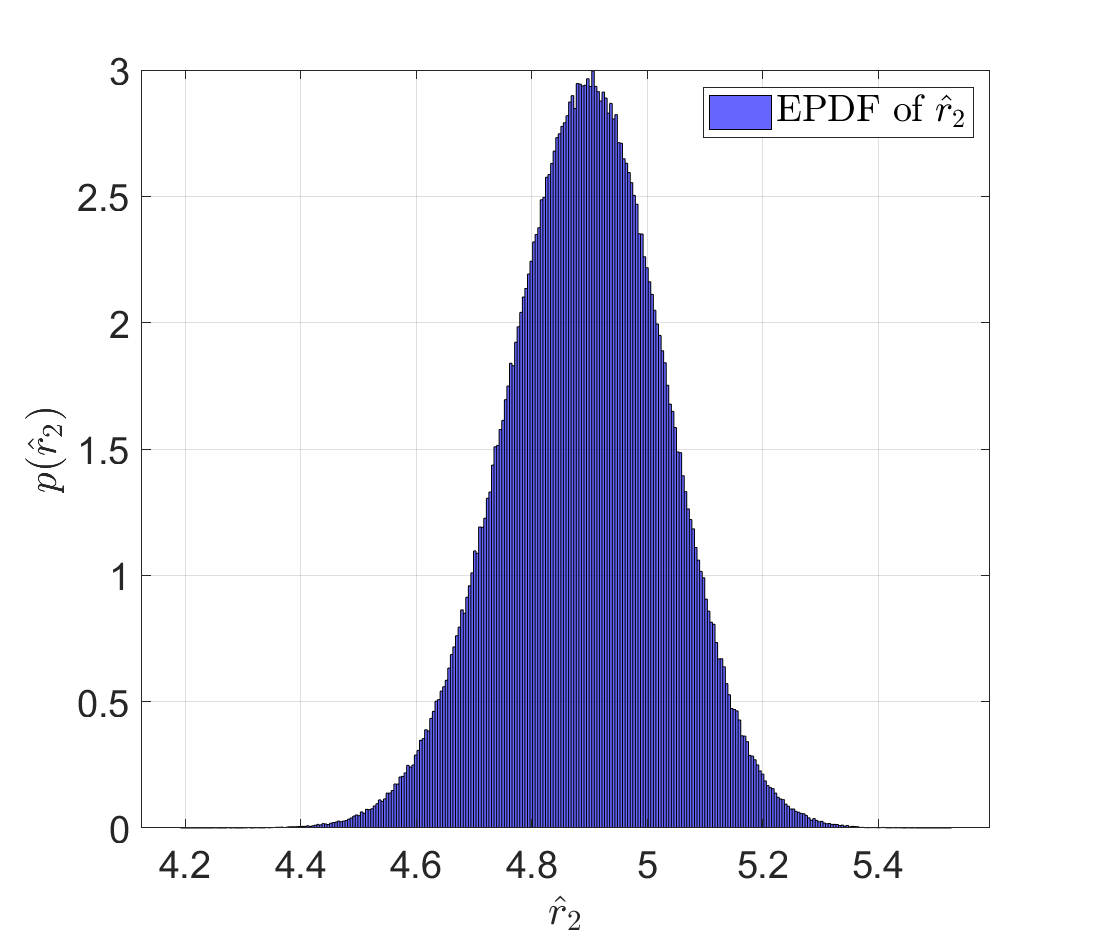}}
\\
\subfloat[EPDF of $\hat{r}_3$]{\includegraphics[height=2.4in]{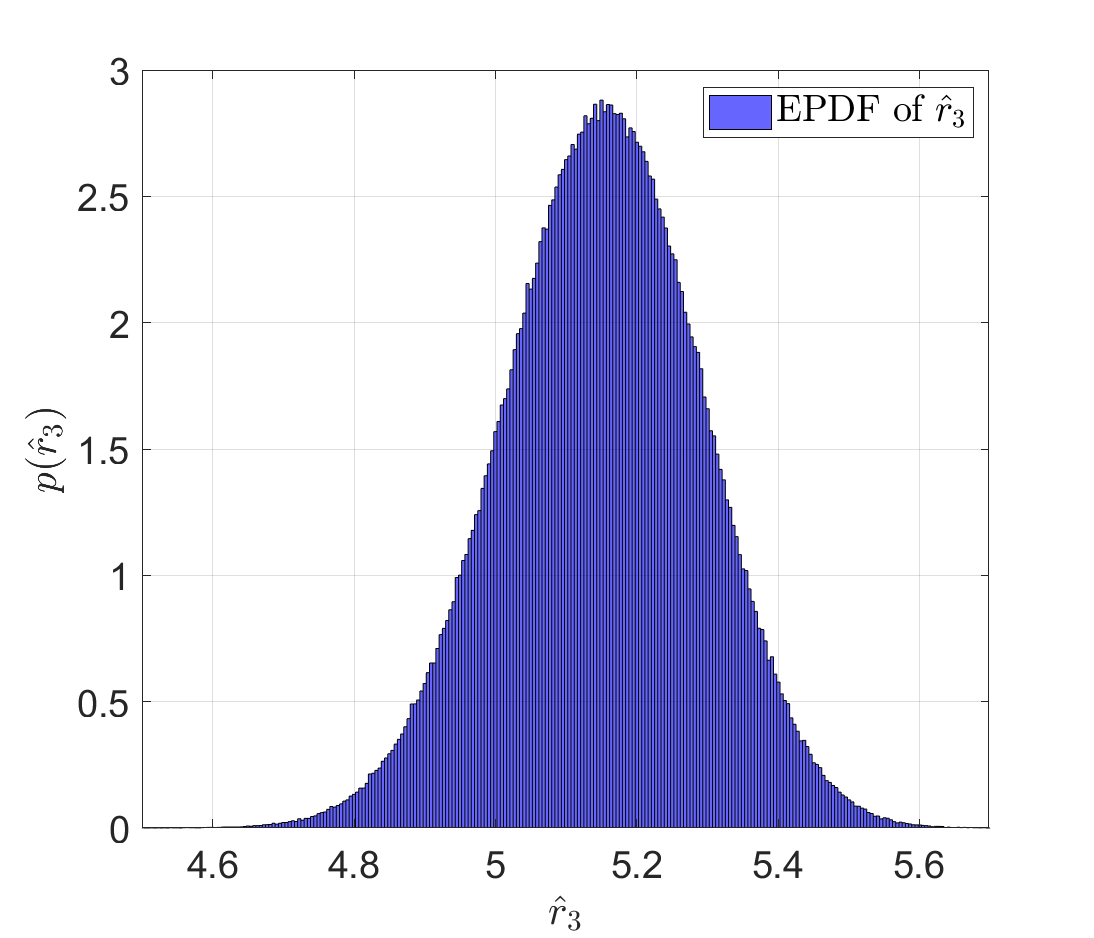}}
\caption{The empirical \ac{pdf} of the radiuses learnt by the designed \ac{nn} at $\gamma=24$ dB for $64$-QAM.}\label{classificationyyyyyyyyyyy}
\end{figure}


Unless otherwise mentioned, $q=3$ is considered, and \ac{mmse} is employed as the suboptimal detector. The performance of the proposed \ac{dl}-based sphere decoding algorithm in terms of \ac{ber} and computational complexity is obtained from
$10^6$ Monte Carlo trials for each \ac{snr} value.

The performance comparison of the \ac{dl}-based sphere decoding algorithm with \ac{spi} algorithm in \cite{vikalo2005sphere} and its Schnorr-Euchner (SE) variate with SE-\ac{spi} in \cite{zhao2005sphere} are performed with the same sets of fading channel matrixes, transmit vectors, and noise vectors. For fair comparison, it is considered that SE-\ac{spi} uses $q$ radiuses, and then switches to MMSE decoding after $q$ times radius increasing. The decoding hypersphere radiuses for \ac{spi} and SE-\ac{spi} were obtained
for $p_{\rm{c}}(i)=1-0.01^i$ at the $i$th sphere decoding implementation as suggested in \cite{vikalo2005sphere}.

\subsection{Simulation Results}
Fig. \ref{classificationyyyyyyyyyyy} shows the underlaying empirical \ac{pdf} of the radiuses learnt by the designed \ac{nn} at $\gamma=24$ dB for $64$-QAM.
As expected, the radiuses vary based on the inputs of the \ac{nn}. This implies that the radiuses are intelligently adjusted to the channel state information and
received signals.

Fig. \ref{classificationyyyyyyyyyyyytber} illustrates the \ac{ber} of the proposed \ac{dl}-based sphere decoding algorithm and its SE variate versus the average \ac{snr}.
As seen, the proposed \ac{dl}-based algorithm exhibits \ac{ber} performance close to that in \ac{spi} (\ac{mld}) over a wide range of \ac{snr}s. This behaviour shows that sequential sphere decoding implementation with the learned radiuses reaches the optimal solution.

\begin{figure}[!t]
\subfloat[64-QAM]{\includegraphics[height=2.8in]{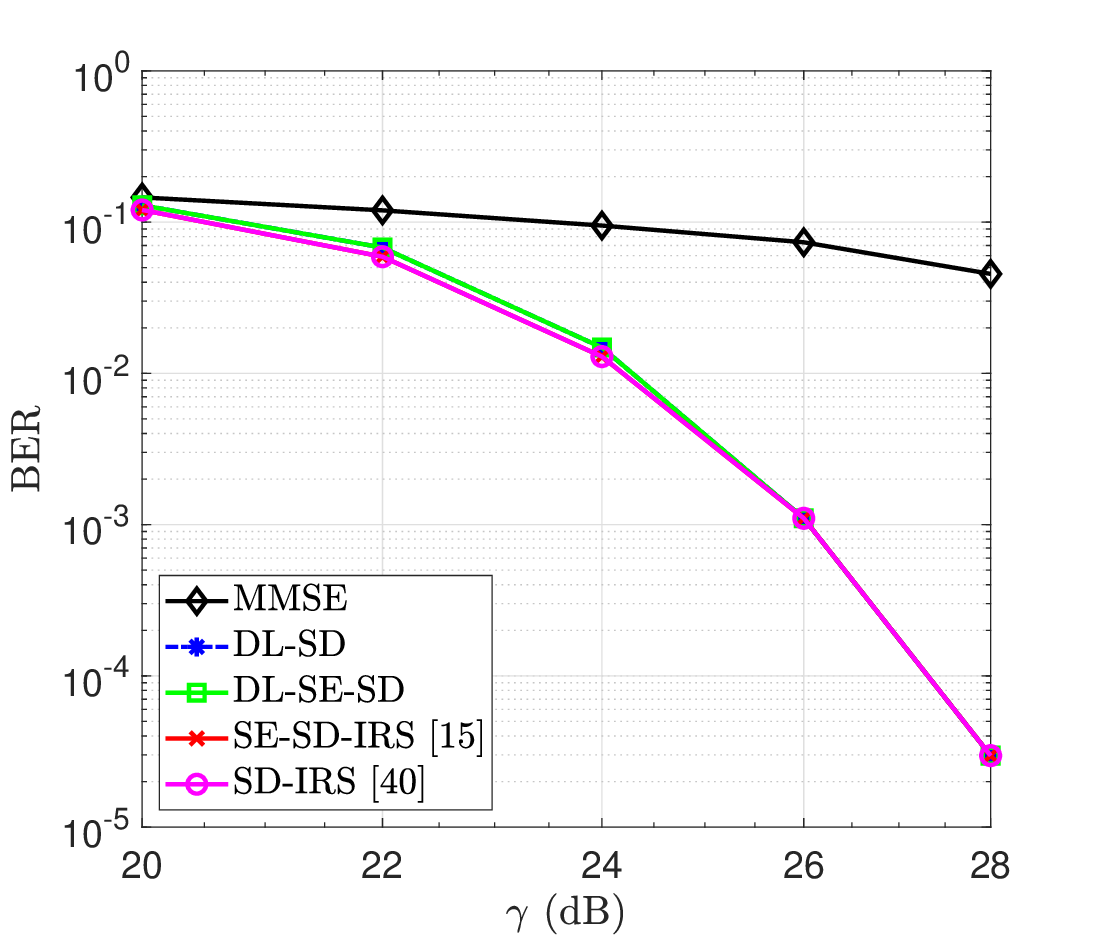}}
\\
\subfloat[16-QAM]{\includegraphics[height=2.8in]{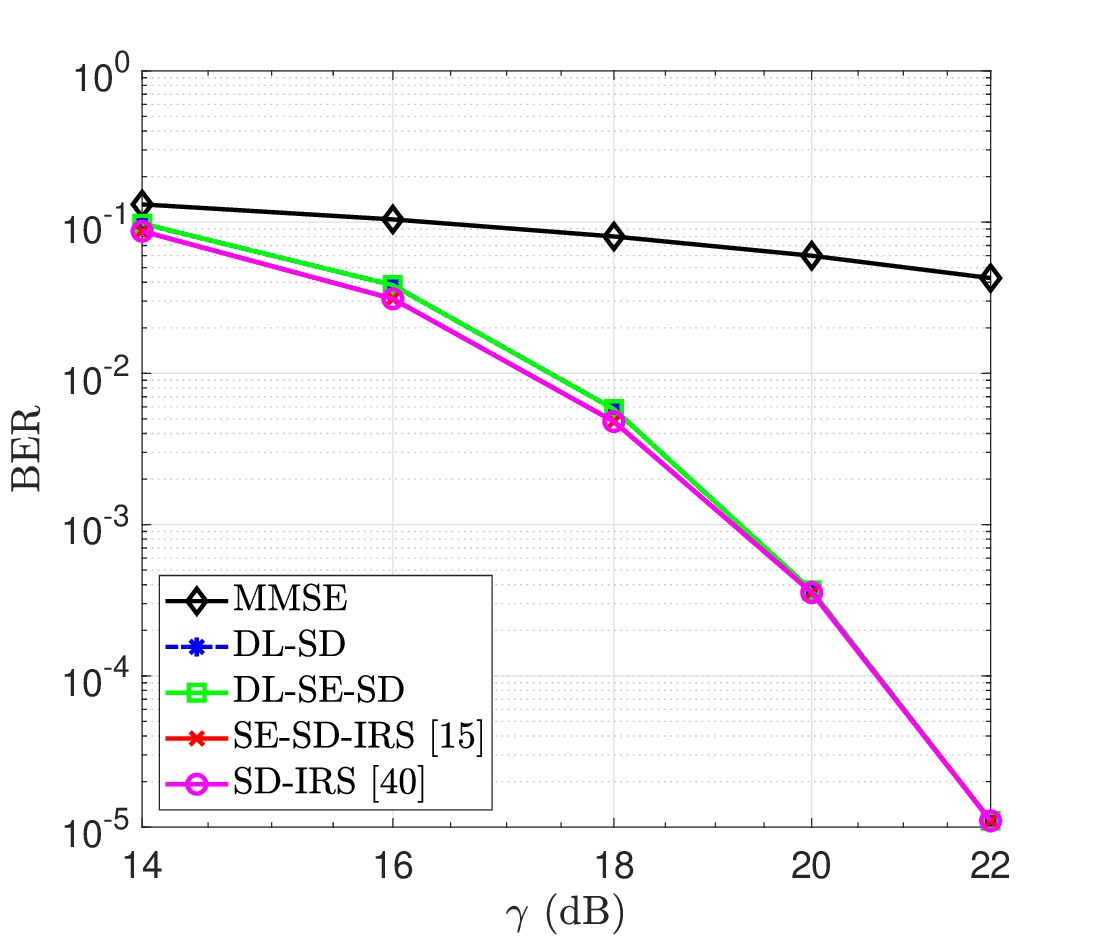}}
\caption{Performance comparison of the proposed \ac{dl}-based sphere decoding algorithm and its SE variate ($q=3$), the \ac{spi} \cite{vikalo2005sphere}, and SE-\ac{spi} in \cite{zhao2005sphere}.
}\label{classificationyyyyyyyyyyyytber}
\end{figure}

\begin{figure*}[t!]
\hspace{1em}
\subfloat[64-QAM]{\includegraphics[height=2.95in]{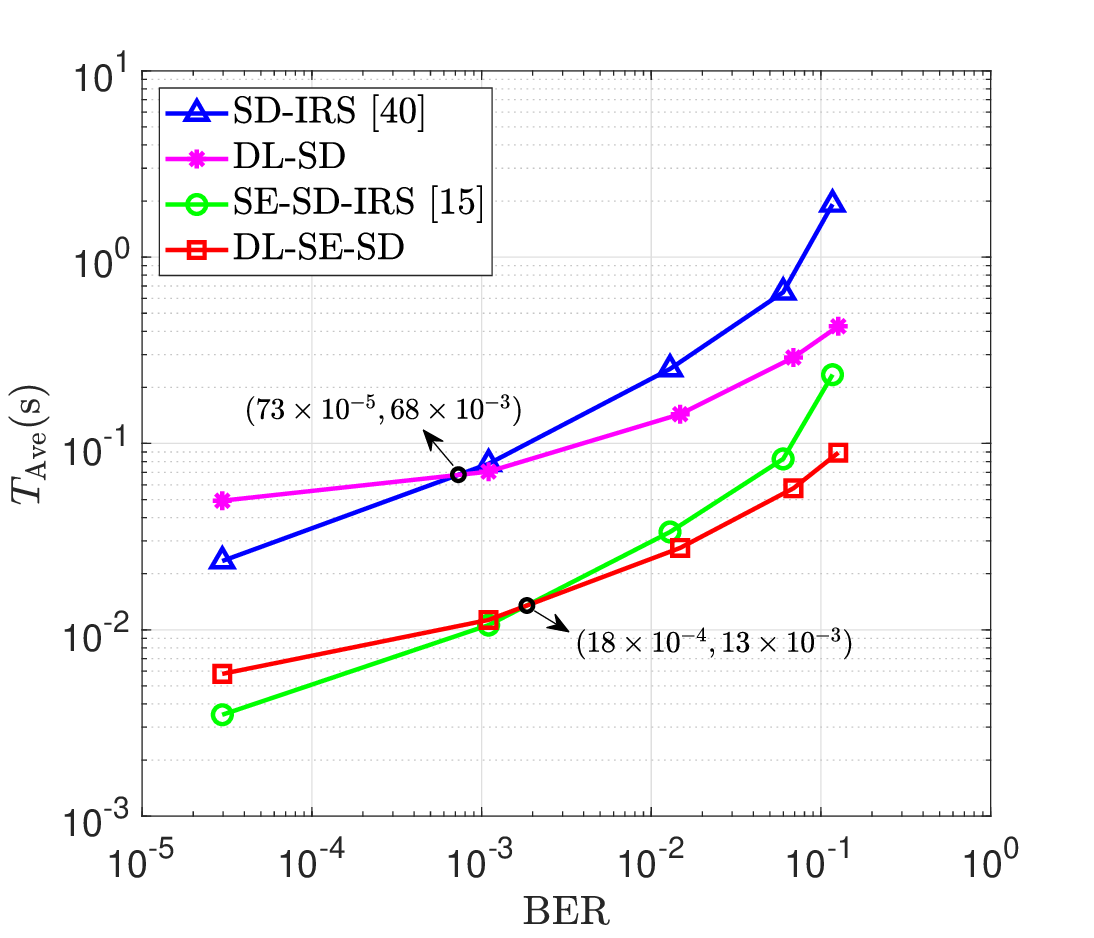}}
\subfloat[16-QAM]{\includegraphics[height=2.95in]{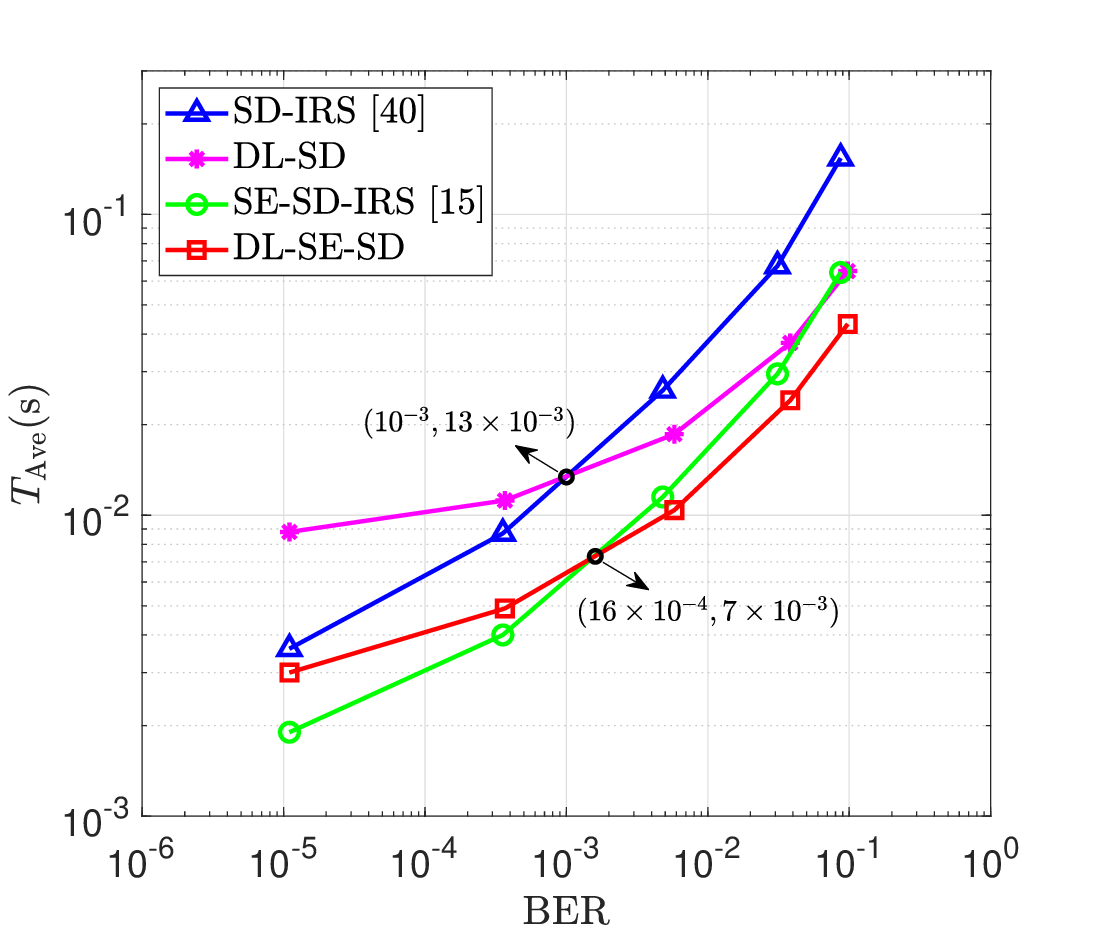}}
\caption{Average decoding time versus BER. The corresponding SNR of markers (left to right) for 64-QAM and 16-QAM are $\{28,26,24,22,20\}$ dB and $\{22,20,18,16,14\}$ dB, respectively.} \label{classificationyyyyyyyyyyyyt6YUYU}
\end{figure*}
\begin{figure*}[t!]
\hspace{1em}
\subfloat[]{\includegraphics[height=2.96in]{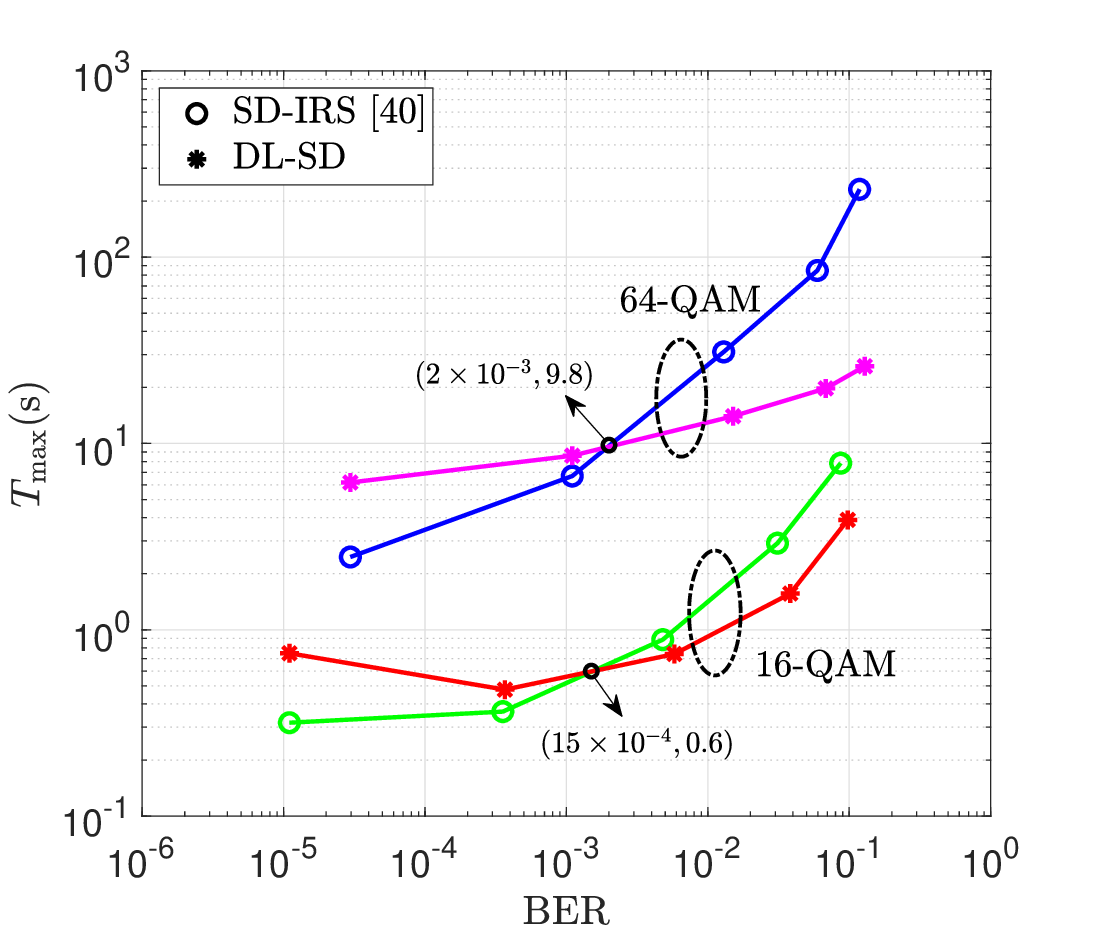}}
\subfloat[]{\includegraphics[height=2.96in]{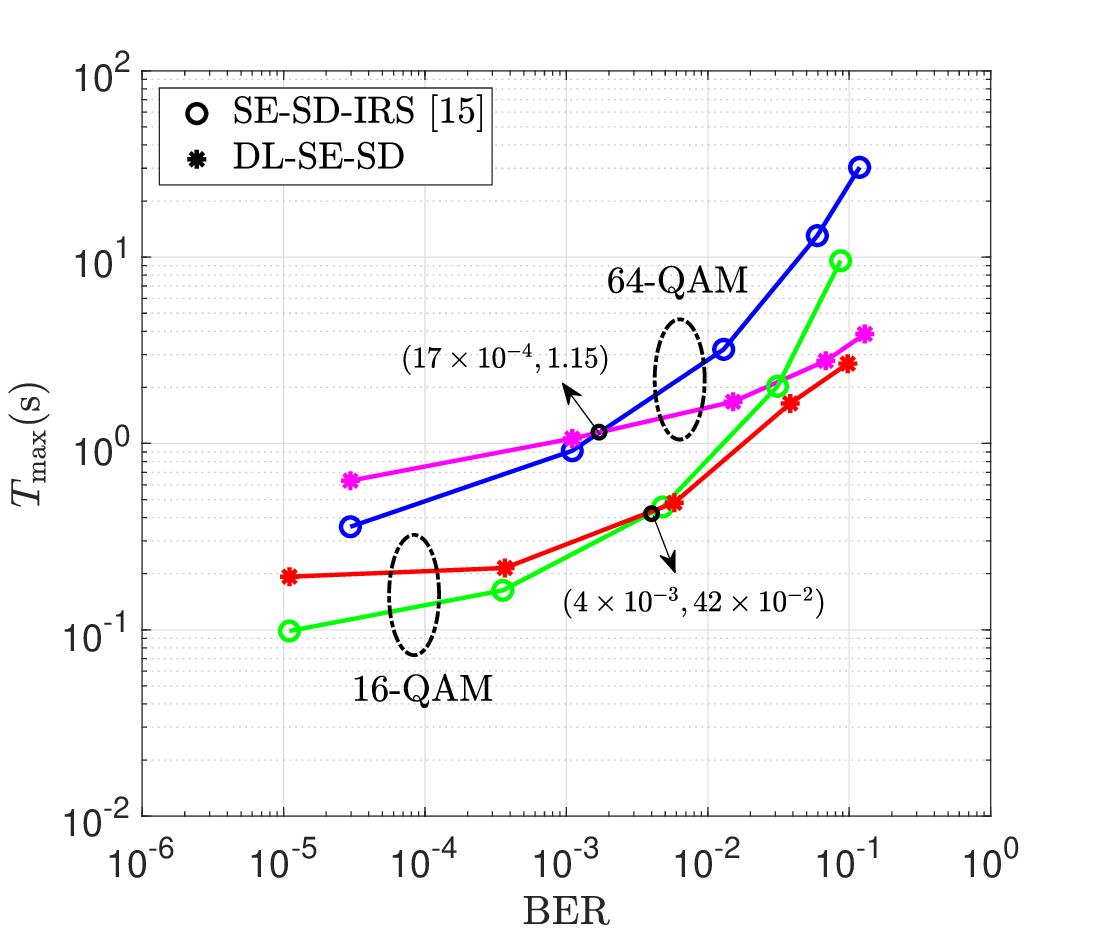}}
\caption{Maximum decoding time versus BER. The corresponding SNR of markers (left to right) for 64-QAM and 16-QAM are $\{28,26,24,22,20\}$ dB and $\{22,20,18,16,14\}$ dB, respectively.}\label{classificationyyyyyyyyyyyyt6maz}
\end{figure*}

Fig. \ref{classificationyyyyyyyyyyyyt6YUYU} illustrates the average decoding time of the proposed \ac{dl}-based sphere decoding algorithm and its SE variate versus \ac{ber}.
As seen, the average decoding time in \ac{dl}-based algorithm
is significantly lower than SD-IRS \cite{vikalo2005sphere} when $ {\text{BER}}> 73\times 10^{-5}$ for $64$-QAM, and when $ {\text{BER}}>   10^{-3}$ for $16$-QAM.
Also, as seen, the SE variate of the proposed \ac{dl}-based algorithm outperforms SE-SD-IRS \cite{zhao2005sphere} when ${\text{BER}}> 18 \times 10^{-4}$ for $64$-QAM, and when $ {\text{BER}}> 16 \times 10^{-4}$ for $16$-QAM.
The reason for this reduction in complexity is that the number of lattice points inside the decoding hypersphere, and thus the
size of the search tree decreases in the average sense when the radiuses of the hyperspheres are intelligently learnt by a \ac{dnn}.
On the other hand, the SD-IRS \cite{vikalo2005sphere} and SE-SD-IRS \cite{zhao2005sphere} exhibit a lower computational complexity compared to the proposed \ac{dl}-based algorithm and its SE variate at very low \ac{ber} regions (high SNRs). The reason is that at high \ac{snr} values, it is unlikely for the lattice to collapse in one or more dimension, an event that significantly increases the number of points in the hypersphere for the scheme in \cite{vikalo2005sphere} and \cite{zhao2005sphere}.
One possible way to improve the proposed method is to consider
\ac{snr}-based \ac{dnn}, especially at high \ac{snr} values.

 \begin{figure}[!t]
\centering
    \includegraphics[height=2.95in]{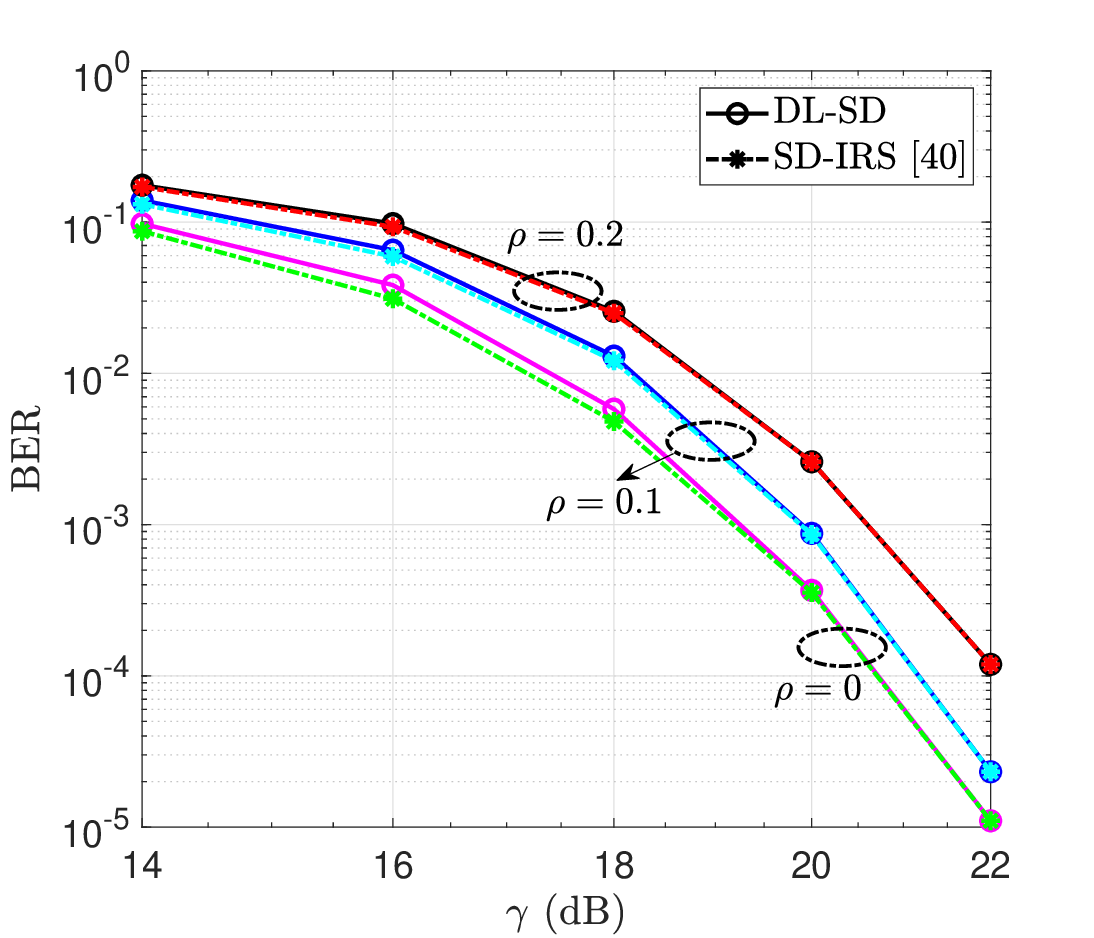}
    \vspace{1em}
    \caption{Performance of the proposed \ac{dl}-based sphere decoding algorithm for $16$-QAM and $q=3$ in the presence of spatial correlation mismatch in the training and decoding phases.  }
    \label{fig:long_shortyussdsdsfsdffrfrfrfr55555552}
\end{figure}

 \begin{figure}[]
\centering
    \includegraphics[height=2.95in]{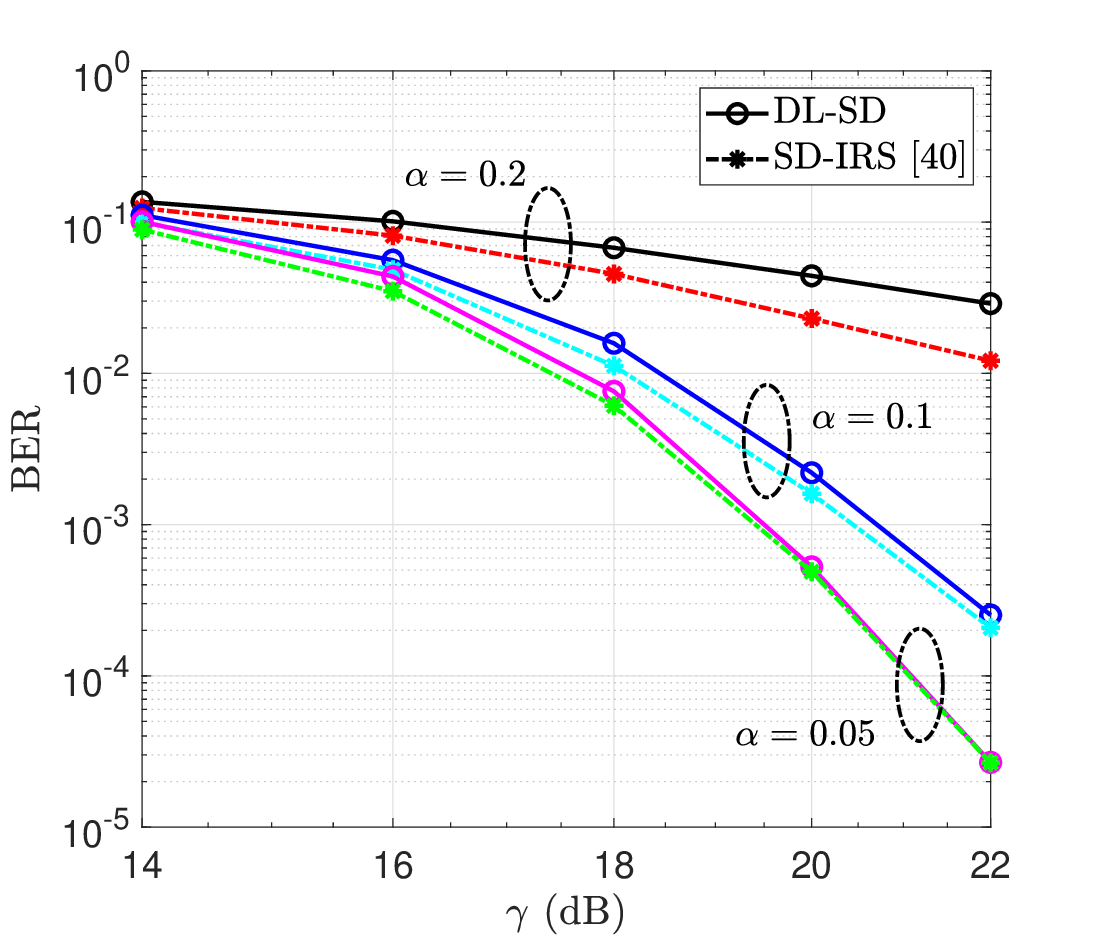}
    \vspace{1em}
    \caption{Performance of the proposed \ac{dl}-based sphere decoding algorithm for $16$-QAM and $q=3$ in the presence of channel estimation error. }
    \label{fig:long_shortyussdsdsfsdffrfrfrfruu}
\end{figure}

\begin{figure*}[]
\subfloat[64-QAM]{\includegraphics[height=2.95in]{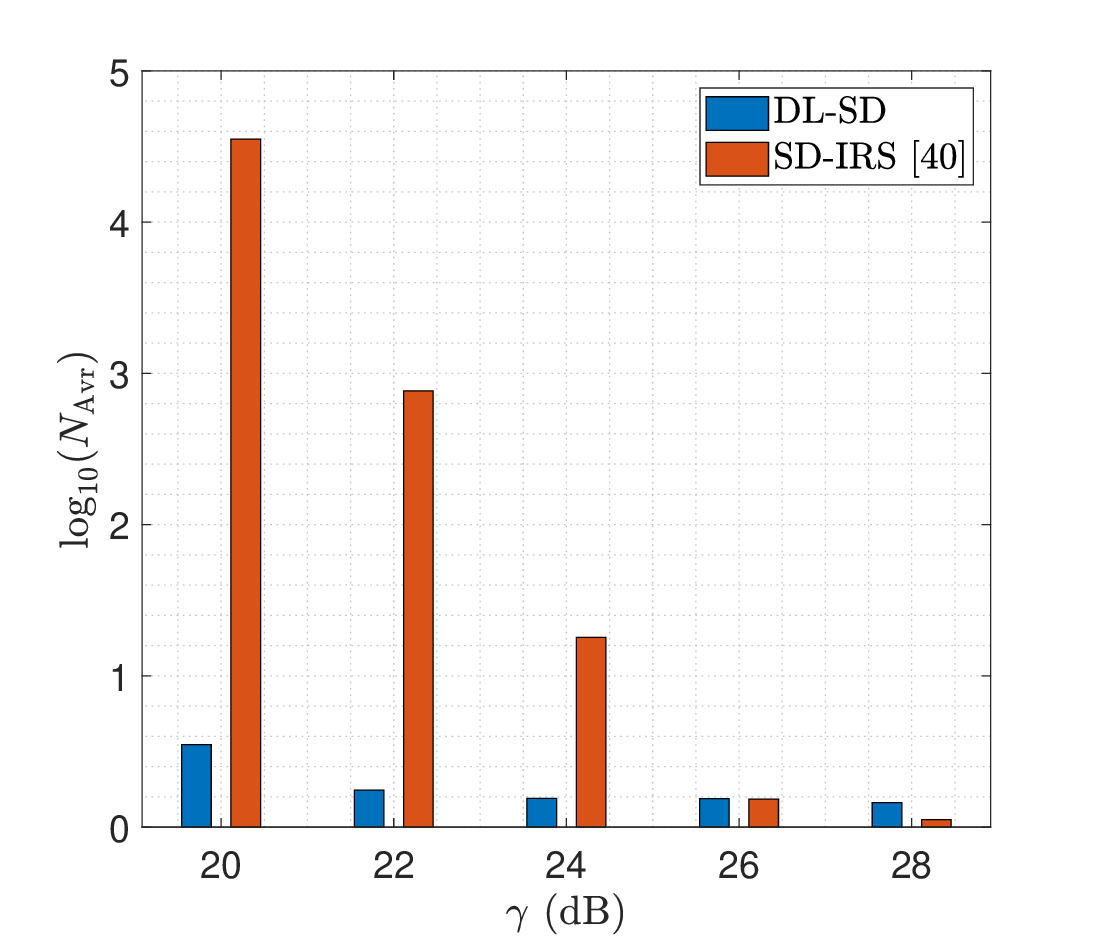}}
\subfloat[16-QAM]{\includegraphics[height=2.95in]{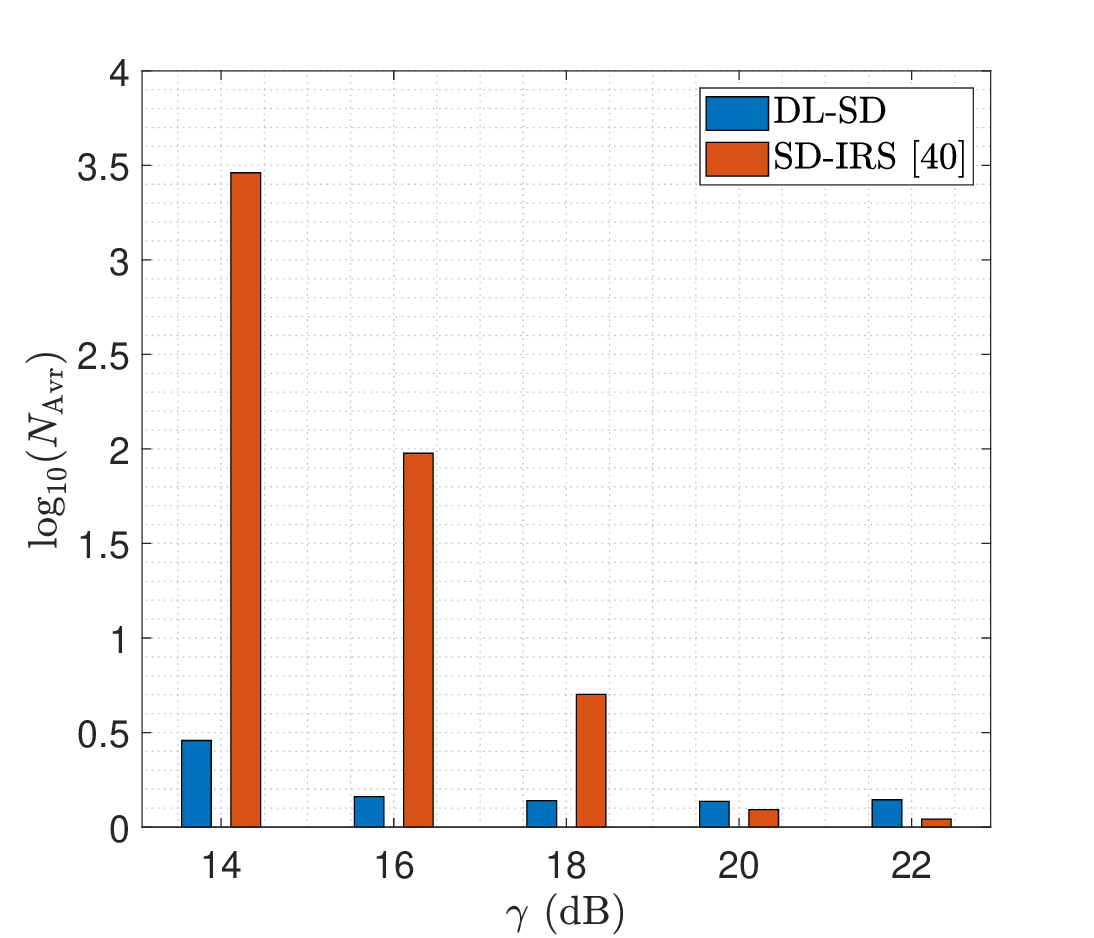}}
\caption{The average number of lattice points (in the logarithmic scale) falling inside the search hypersphere in the \ac{dl}-based sphere decoding algorithm and the \ac{spi} algorithm in \cite{vikalo2005sphere}.}\label{classificationyyyyyyyyyyyyt23}
\end{figure*}

Fig. \ref{classificationyyyyyyyyyyyyt6maz} shows the maximum decoding time in the proposed \ac{dl}-based algorithm and its SE variate versus BER to that in the \ac{spi} algorithm \cite{vikalo2005sphere} and SE-SD-IRS \cite{zhao2005sphere}.
As seen, the \ac{dl}-based sphere decoding algorithms outperform the algorithm in \cite{vikalo2005sphere} and \cite{zhao2005sphere} for some regions, i.e., BER  $>2 \times 10^{-3}$ for DL-SD and BER  $>17\times 10^{-4}$ for DL-SE-SD when using $64$-QAM.
These regions for $16$-QAM are BER  $> 15 \times 10^{-4}$ and BER  $> 4 \times 10^{-3}$, respectively.
This shows that the size of the search tree in the \ac{dl}-based sphere decoding is much smaller
than the one in the algorithm in \cite{vikalo2005sphere} and \cite{zhao2005sphere} in the worst-case sense.


While at lower BERs (higher SNRs) \ac{spi} and SE-\ac{spi} offer a lower complexity compared to our proposed solutions, one should note that the presented figures represents \ac{ber} in the absence of channel coding. In practice, when lower BERs are needed, channel coding is used. Hence, for the practical range of moderate BER before channel coding, our proposed solution offers a better complexity.

Fig. \ref{fig:long_shortyussdsdsfsdffrfrfrfr55555552} shows the \ac{ber} of the proposed \ac{dl}-based algorithm for $16$-QAM and $q=3$ in the presence of distribution mismatch. It is assumed that the \ac{dnn} is trained for independent fading channel; however, it is evaluated in the presence of correlated fading channel. As seen, the proposed algorithm is robust to spatial correlation fading for $\rho = 0.1$ and $\rho = 0.2$, where $\rho$ is the complex correlation coefficient of neighboring transmit and receive antennas.

Fig. \ref{fig:long_shortyussdsdsfsdffrfrfrfruu} shows the \ac{ber} of the proposed \ac{dl}-based algorithm for $16$-QAM and $q=3$ in the presence of channel estimation error.
It is assumed that the real and imaginary parts of the estimated fading channel $\hat{h}_{ij}$ are randomly drawn form uniform distribution as
 \begin{align} \nonumber
  &\Re \{\hat{h}_{ij}\} \in \mathcal{U}\Big{(} (1-\alpha) \Re \{{h}_{ij}\} , (1+\alpha) \Re \{{h}_{ij}\} \Big{)} \\ \nonumber
  &\Im \{\hat{h}_{ij}\} \in \mathcal{U}\Big{(} (1-\alpha)  \Im \{{h}_{ij}\} , (1+\alpha)  \Im \{{h}_{ij}\}\Big{)},
\end{align}
where
${h}_{ij}$ is the true value of channel, and $\alpha$ is a parameter used to control channel estimation error.
The effect of channel estimation error on the \ac{ber} for three values of $\alpha \in \{0.05,0.1,0.2\}$ is shown in Fig. \ref{fig:long_shortyussdsdsfsdffrfrfrfruu}.
As expected, the performance of sphere decoding (DL and IRS) depends on the accuracy of channel estimation. Thus, a lower $\alpha$ results in a lower performance degradation. In addition, as seen, the performance of the proposed DL-based algorithm is robust to the channel estimation error when $\alpha<0.1$.

In Fig. \ref{classificationyyyyyyyyyyyyt23}, the average number of lattice points (in the logarithmic scale) falling inside the decoding hypersphere in the \ac{dl}-based sphere decoding algorithm is compared with the one in the \ac{spi} algorithm in \cite{vikalo2005sphere}.
As seen, the average number of lattice points in the \ac{dl}-based algorithms is below $0.545$ (in the non logarithmic scale, bellow $3.51$),  while this is much higher in the \ac{spi} algorithm.


\section{Conclusion}\label{5629479274-2852}
A low-complexity solution for integer \ac{ls} problems based on the capabilities of \ac{dl} and sphere decoding algorithm was proposed in this paper. The proposed solution leads to efficient implementation of sphere decoding for a small set of intelligently learned radiuses.
The \ac{ber} performance of the \ac{dl}-based sphere decoding algorithm is very close to that in \ac{mld} for high-dimensional integer \ac{ls} problems with significantly lower computational complexity.
The expected complexity of the proposed algorithm based on the elementary operations was derived, and its effectiveness in term of \ac{ber} and computational complexity for high-dimensional \ac{mimo} communication systems, using higher-order modulations,  was shown through simulation.
While the integer \ac{ls} problem in this paper was formulated for \ac{mimo} communication systems, it is a promising solution for other situations when integer \ac{ls} problems are encountered, such as multi-user communications,  relay communications, global positioning system, and more.

{\section*{Acknowledgements}
The authors are grateful to the anonymous reviewers and the
Editor, Dr. Xiangyun Zhou, for their constructive comments.}

\appendices
\section{}\label{ryryrrz7323}
The expected complexity of sphere decoding implementation for radius $d$ is given as \cite{vikalo2005sphere}
\begin{align}\label{tyuhgge12}
C(m,\sigma_{\rm{w}}^2,d)=\sum_{k=1}^{m}F_{\rm{sp}}(k) \sum_{v=0}^{\infty}\gamma \Big{(}\frac{d^2}{\sigma_{\rm{w}}^2+v},n-m+k\Big{)}\Psi_{2k}(v).
\end{align}

By employing \eqref{tyuhgge12} and following the same procedure as in \cite{vikalo2005sphere}, the expected complexity of the \ac{spi} algorithm for $r_1 < r_2 < \cdots < r_q$ is obtained  as
\begin{align}\label{yyyy7tttrtyi}
C(m,\sigma_{\rm{w}}^2,r_1,\cdots,r_q)=&\sum_{c=1}^{q}(p_c-p_{c-1})\sum_{k=1}^{m}F_{\rm{sp}}(k) \\ \nonumber
&\hspace{-0.5em}\times \sum_{v=0}^{\infty}\gamma \Big{(}\frac{r_{c}^2}{\sigma_{\rm{w}}^2+v},n-m+k\Big{)}\Psi_{2k}(v),
\end{align}
where $p_0=0$, and $p_c$, $0<c \leq q$, is the probability of finding at least a lattice point inside the hypersphere with radius $r_c$, which is obtained by replacing  $r_c$ with $\hat{r}_{i_c}$ in \eqref{tgdjgadgagdjh67661w19721}.

The probability that a solution is not found during the sphere decoding implementation for the hypeespheres with radiuses $r_1, r_2, \cdots, r_q$ equals $(1-p_q)$. Hence, the proposed \ac{dl}-based sphere decoding algorithm obtains the solution through a suboptimal detector  with probability $(1-p_q)$. This leads to $(1-p_q)F_{\rm{sb}}$ additional average complexity given a suboptimal detector with $F_{\rm{sb}}$ elementary operations.
For the \ac{mmse} detector in \eqref{88fjbbk785rrrrttyy},
the number of elementary operations of $({\bf{H}}^{\rm{H}}{\bf{H}}+\bar{\gamma}^{-1}{\bf{I}})$ is $nm^2+m(n-\frac{m}{2})+\frac{m}{2}$, the matrix inversion in \eqref{88fjbbk785rrrrttyy} requires $m^3+m^2+m$ elementary operations, ${\bf{H}}^{\rm{H}}\bf{y}$ requires $m(2n-1)$ elementary operations, and the product of $({\bf{H}}^{\rm{H}}{\bf{H}}+\bar{\gamma}^{-1}{\bf{I}})^{-1}$ and ${\bf{H}}^{\rm{H}}\bf{y}$ requires $2m^2-m$ elementary operations \cite{golub2012matrix}. Thus, the total elementary operations in the \ac{mmse} detection is given as in  \eqref{5557880785qasx}.

Moreover, there is $F_{\rm{dn}}$ elementary operations due to the \ac{dnn} computations.  The number of multiplication and addition in the $\ell$th layer of a \ac{dnn} with $n_{\ell}$ neurons is $2n_{\ell} n_{\ell-1}$, where $n_{\ell-1}$ is the number of neurons in the $(\ell-1)$th layer.  Hence, for a $L$-layer \ac{dnn} with $n_0, \cdots n_{L}$ neurons in each layer, $F_{\rm{dn}}$ is given as in \eqref{ghssdkhckshdkh3}.

By employing \eqref{yyyy7tttrtyi} and including $(1-p_q)F_{\rm{sb}}$ and $F_{\rm{dn}}$,
the expected complexity of the proposed \ac{dl}-based sphere decoding algorithm given the learned radiuses $\hat{r}_{i_1}, \cdots, \hat{r}_{i_q}$ is obtained as
\begin{align}\nonumber
&C_{\rm{DL}}(m,\sigma_{\rm{w}}^2,\hat{r}_{i_1},\cdots,\hat{r}_{i_q})=\sum_{c=1}^{q}(\hat{p}_{i_c}-\hat{p}_{{i_{c-1}}})\sum_{k=1}^{m}F_{\rm{sp}}(k) \\
&\times \sum_{v=0}^{\infty}\gamma \Big{(}\frac{\hat{r}_{i_c}^2}{\sigma_{\rm{w}}^2+v},n-m+k\Big{)}\Psi_{2k}(v)+(1-\hat{p}_{i_q})F_{\rm{sb}}+F_{\rm{dn}},
\end{align}
where $\hat{p}_{i_c}$ is given in \eqref{tgdjgadgagdjh67661w19721}. Finally,
since $\hat{r}_{i_1}, \cdots ,\hat{r}_{i_q}$ and thus $\hat{p}_{i_1}, \cdots ,\hat{p}_{i_q}$ are random variables, one can write the expected complexity of the \ac{dl}-based sphere decoding as in \eqref{hhhbbccxzzrr56577}.

\end{document}